\RequirePackage{fix-cm}
\documentclass[fleqn,usenatbib]{mnras}

\usepackage{newtxtext,newtxmath}
\usepackage[T1]{fontenc}
\DeclareRobustCommand{\VAN}[3]{#2}
\let\VANthebibliography\thebibliography
\def\thebibliography{\DeclareRobustCommand{\VAN}[3]{##3}\VANthebibliography}

\usepackage{graphicx}	
\usepackage{amsmath}	
\DeclareUnicodeCharacter{2212}{-}

\usepackage{booktabs}
\usepackage{longtable} 
\usepackage{multirow}
\usepackage{adjustbox}  % Add in preamble
\usepackage{makecell}

\title[Thermodynamics of ICMEs at 1 AU]{Thermal properties of interplanetary coronal mass ejections at 1 AU and their connection to geoeffectiveness across solar cycles 23–25}

\author[S. Khuntia et al.]{Soumyaranjan Khuntia \orcid{0009-0006-3209-658X},$^{1, 2}$\thanks{E-mail: soumyaranjan.khuntia@iiap.res.in (SK)}
 and Wageesh Mishra \orcid{0000-0003-2740-2280},$^{1, 2}$\thanks{E-mail: m.wageesh30@gmail.com (WM)}
\\
$^{1}$Indian Institute of Astrophysics, II Block, Koramangala, Bengaluru 560034, India\\
$^{2}$Pondicherry University, R.V. Nagar, Kalapet 605014, Puducherry, India
}

\pubyear{2024}

\begin{document}
\label{firstpage}
\pagerange{\pageref{firstpage}--\pageref{lastpage}}
\maketitle

\begin{abstract}
Interplanetary coronal mass ejections (ICMEs) are major drivers of heliospheric variability and can produce prolonged disturbances near Earth. Understanding their thermodynamic evolution is crucial for assessing their heat budget and exploring how thermal states relate to their plasma dynamics and geoeffectiveness. We conduct a comprehensive statistical analysis of magnetic ejecta (MEs) over Solar Cycles 23, 24, and the ascending phase of 25. Leveraging a polytropic framework, we characterized the thermal state of ME based on the event-wise median proton polytropic index ($\Gamma_p$) from in-situ measurements at 1 AU. We find that MEs are thermodynamically active and rarely evolve adiabatically or isothermally. Notably, a significant fraction (45\%) of MEs exhibit a heating state. Heating MEs dominate near solar maxima and exhibit strong solar-cycle modulation in $\Gamma_p$, proton temperature, and expansion speed, indicating active in-transit heating processes. Whereas, Cooling MEs show a nearly constant $\Gamma_p \sim 2$ across cycles, suggesting enhanced cooling beyond adiabatic expectations and possible thermal energy retention from eruption to 1 AU. Notably, the median $\Gamma_p$ value increases from 1.49 (SC23) to 1.88 (SC24), indicating a shift to cooling-dominated states over successive cycles. High-impact ICMEs, predominantly Heating MEs ($\Gamma_p = 0.59$), often manifest as magnetic clouds with enhanced magnetic fields, low plasma beta, pronounced sheath compression, elevated expansion, and post-ICME high-speed flows, making them the most geoeffective drivers of strong geomagnetic storms. These results establish $\Gamma_p$ as a useful diagnostic of ICME thermal states, though meaningful assessment of geoeffectiveness requires combined consideration of thermal, plasma, and magnetic field properties. 
\end{abstract}


\section{Introduction}

Coronal mass ejections (CMEs) are large-scale eruptions of magnetized plasma from the Sun’s corona that propagate through the heliosphere, often reaching Earth’s orbit (near 1 AU) and playing a crucial role in driving space weather disturbances \citep{Gosling1993, Tsurutani1988, Webb2012, Temmer2023}. CMEs are commonly observed close to the Sun using remote sensing instruments such as coronagraphs and heliospheric imagers. These instruments detect photospheric white light that is scattered by free electrons in the CME plasma via Thomson scattering. In addition to this, in-situ measurements from space-based probes provide valuable complementary information as CMEs propagate through the heliosphere. Once CMEs propagate through the interplanetary medium, they are traditionally referred to as interplanetary coronal mass ejections (ICMEs). A significant link between CMEs and ICMEs was established in 1982 when an ICME detected by Helios-1 was associated with a white-light CME \citep{Burlaga1982}. During their propagation, ICMEs interact with the ambient solar wind (SW) and the interplanetary magnetic field (IMF), and upon reaching Earth, they have potential of triggering major geomagnetic storms (GS). The arrival of an ICME is typically preceded by a shock and an associated sheath region (compressed SW plasma). The shock and sheath is often followed by an ICME having a prolonged period of southward-oriented magnetic field ($B_z$), which facilitates magnetic reconnection with Earth’s northward-pointing field lines \citep{Gopalswamy2006, Zurbuchen2006, Zhang2007, Echer2008}. This reconnection is the primary mechanism for transferring energy from ICME into the magnetosphere, ultimately driving geomagnetic disturbance \citep{Gonzalez1994}. Due to their significant impacts on satellite systems, communication networks, power infrastructure, and many other technological systems \citep{Pulkkinen2007, Baker2009, Temmer2021}, the association between CMEs properties near the Sun and at 1 AU have become a major focus in space weather research \citep{Liu2010,Mishra2013,Chi2016,Mostl2022}. The fraction of time Earth remains within ICMEs is approximately 10\% during solar minimum, increasing to as much as 35\% during solar maximum \citep{Cliver2001}. Gaining a comprehensive understanding of their kinematic, magnetic, and thermal properties is essential for improving space weather forecasting capabilities.

The identification of ICMEs is generally based on characteristic variations in several properties of the magnetized plasma \citep{Zurbuchen2006, Wu2011}. A key component of an ICME is the magnetic ejecta (ME), characterized by a magnetic field that is stronger than the ambient SW and often exhibits smooth rotation \citep{Burlaga1982}; low plasma beta (the ratio of thermal pressure to magnetic pressure) and low proton temperatures \citep{Richardson1995}; low electron temperatures \citep{Montgomery1974}; abnormal high charge state of ions \citep{Lepri2001}; and bidirectional streaming of electrons \citep{Gosling1987}. In addition, the presence of a declining flow speed profile, indicative of the expansion of ICMEs, can also be considered a key signature for their identification \citep{Russell2003, Zurbuchen2006, Jian2006}. However, it is important to note that none of these signatures alone is unique to ICMEs or sufficient for definitive identification \citep{Zurbuchen2006,Mishra2023}. Magnetic clouds (MCs) represent a distinct subset of ME, characterized by low plasma beta and a smoothly rotating coherent magnetic field spanning a large angle, $\approx 180^{\circ}$ \citep{Burlaga1981, Burlaga1982}. The multi-spacecraft studies have demonstrated that the detection of such MC structures strongly depends on the spacecraft’s trajectory through the ICME \citep{Cane1997, Kilpua2011}. If the spacecraft passes far from the central axis or along the legs of the structure, the characteristic twisted magnetic configuration may not be detected, leading to the non-identification of a flux rope even if one is present. It is estimated that approximately 30-50\% of ICMEs observed at 1 AU exhibit MC signatures that may have some variations over solar cycles \citet{Gosling1990, Cane1997, Wu2007,Mishra2021a}. As ICMEs evolve through interplanetary space, they often undergo interactions with the ambient solar wind \citep{Dasso2006, Vrsnak2010, Ruffenach2012} or with other CMEs leading to change in their kinematics and structures \citep{Lugaz2014, Mishra2014, Mishra2017, Scolini2020}.

The SOlar and Heliospheric Observatory (SOHO: \citealt{Domingo1995}), positioned at the first Lagrange point (L1), continuously tracks CMEs since 1995. Complementing this, the Solar TErrestrial RElations Observatory (STEREO: \citealt{Kaiser2005}), with its heliocentric orbit, provides a unique 3D view of CME propagation. In addition to their remote sensing capabilities, both spacecraft are equipped with in-situ instruments, enabling multi-point measurements of CMEs as they travel through the heliosphere. Interplanetary probes such as the Advanced Composition Explorer (ACE: \citealt{Stone1998}) and Wind \citep{Lepping1995, Ogilvie1995} have been instrumental in continuously monitoring the solar wind and interplanetary magnetic field near Earth for over 25 years. Several reliable ICME catalogs have been developed based on these in-situ measurements from ACE and Wind (\citealt{Richardson2010}: hereafter RC catalog), Wind alone \citep{Chi2016, Nieves2018}, and STEREO \citep{Jian2018}. These catalogs include ME boundaries for each ICME, serving as valuable references for large-sample statistical studies.

One of the major goals in the CME research field is to forecast their arrival time at Earth and assess their potential impact in driving the geomagnetic storms. The magnetic field, speed, and duration/radial sizes of CMEs have been investigated in earlier studies as they are known to be important parameters for driving geomagnetic storms \citep{Srivastava2004, Jian2006, Zhang2007,Echer2008}. Since CMEs are composed of magnetized plasma, their global dynamics, including acceleration and expansion, are certainly influenced by their internal thermodynamic properties. The thermal state of ICMEs at 1 AU may not directly influence their geoeffective potential, it could serve as indicators of other underlying properties responsible for driving strong storms. A better understanding of the additional plasma properties is therefore essential for improving both arrival-time predictions and the assessment of geo-effectiveness. It is required to understand if the thermal properties of CMEs close to the Sun or at 1 AU have any dependency on the magnetic storm characteristics. However, our understanding of the evolution of a CME’s internal thermodynamics remains limited, partly due to observational constraints. While spectroscopic observations can offer thermal diagnostics, they are generally confined to regions near the Sun and do not capture the CME’s full evolution \citep{Antonucci1997, Bemporad2010}. In contrast, in-situ observations, which can give a better measure of the thermal state of ICMEs, are only available at discrete heliocentric distances \citep{Zurbuchen2006, Richardson2010}. Moreover, the sparse spatial distribution of spacecraft limits our ability to observe the same CME at multiple distances simultaneously \citep{Phillips1995, Winslow2021}.

To address these limitations, studies by \citet{Wang2009, Mishra2023, Khuntia2023, Khuntia2024} have utilized CME's 3D kinematics and the flux rope internal state (FRIS) model to derive plasma properties, particularly the thermal state at heights (2-25 R$_\odot$) where no plasma measurements are available. These studies reveal that CMEs undergo multiple heat transfer phases, initially in a heat-releasing state near the Sun, transitioning into a near-isothermal heating phase between 3–9 $R_{\odot}$. This thermal evolution was modeled under the consideration of polytropic behavior, where the pressure $p$ (temperature $T$) and density $\rho$ follow the relation $p \propto \rho^\Gamma$ ($T \propto \rho^{\Gamma-1}$), and $\Gamma$ is the polytropic index. This approach provides a simplified yet insightful representation of CME's heating and cooling without involving the complexity of energy equation modeling. The choice of $\gamma$ is crucial not only for characterizing CME plasma but also for understanding the broader coronal, solar wind dynamics and heat transport in the heliosphere \citep{Riley2003, Manchester2004, Desai2020, Cao2013, Liu2022polymhd, Kuzma2023, Cai2025}. In many numerical studies, a fixed non-adiabatic value of $\gamma$ is assumed, and it has been shown that variations in $\gamma$ can significantly alter the simulated evolution of the solar wind and CMEs \citep{Manchester2004, Wu2004, Liou2014, Mayank2022, Liu2022polymhd}. Empirical estimates of the polytropic index within ICMEs typically range from 1.15 to 1.33 \citep{Osherovich1993, Liu2005, Liu2006}, suggesting substantial local heating. Recently, \citet{Khuntia2025b} reports that both fast and slow ICMEs at 1 AU exhibit near-isothermal expansion, sustained internal heating, and distinct turbulence properties, highlighting the role of Alfvenic fluctuations and intermittent structures in their evolution and local energy dissipation. Additional evidence for CME heating comes from both spectroscopic measurements near the Sun \citep{Filippov2002, Lee2017, Reva2023} and the analysis of ion charge states at 1 AU \citep{Rakowski2007, Lepri2012}. More recently, \citet{Khuntia2025a} employed the FRIS model combined with in-situ data at 1 AU to investigate the thermal characteristics of interacting CMEs which caused a great geomagnetic storm. Their results revealed enhanced electron cooling and a dominant heating trend in protons, suggesting complex thermodynamic interactions during CME-CME encounters. Although several studies have probed the polytropic behavior of CMEs at specific distances or time intervals, a systematic and comprehensive investigation of the thermal state of both ICMEs and the surrounding ambient solar wind at 1 AU in connection to their geo-effective parameters is still limited. This motivates the present study, which aims to statistically characterize the thermal properties of ICMEs at 1 AU using in-situ measurements.

Previous studies have investigated the geoeffectiveness of ICMEs and have established a dependency on key ICME properties such as $B_z$, electric field ($E_y = VB_z$), and solar wind dynamics pressure, both of which are crucial drivers of geomagnetic activity \citep{Echer2008, Richardson2011, Zhao2021}. The studies have also looked into the role of different drivers of geomagnetic storms in governing the rate, strength, growth, and recovery of storms \citep{Echer2008,Richardson2012,Mishra2024}. Furthermore, ion charge states serve as valuable indicators for distinguishing between different types of solar wind and ICMEs, which may have varying probabilities of driving strong geomagnetic storms, even though these charge states are not causal parameters determining storm intensity \citep{Henke1998,Lepri2001,Owens2018}. This implies that ion charge states linked with the thermal state of solar wind/ICMEs close to the Sun may have an indirect relationship with their geoeffectiveness. The role of the ICME’s internal thermodynamic state near the Earth in modulating geomagnetic storm intensity has received relatively less attention. Given that we are statistically analyzing the thermal states of ICMEs, it becomes particularly relevant to examine whether CMEs associated with extreme geomagnetic storms exhibit distinct thermal characteristics compared to the broader ICME population. Since the thermodynamic evolution of a CME governs its expansion, density, and pressure gradients, it may also indirectly represent the CME's ability to efficiently transfer energy to Earth's magnetosphere and drive strong geomagnetic responses.  Therefore, studying the thermal behavior of both the CME and the surrounding solar wind at 1 AU could offer additional insights, together with magnetic field, speed, and density, into the extreme space weather.

In Section \ref{sec:data}, we describe the dataset used for our comprehensive analysis. Section \ref{sec:icme_no} details the annual occurrence of different ME categories over solar cycles 23-25. In Section \ref{sec:icme_thermal}, we focus on the estimation of the thermal state of ICMEs at 1 AU and explore its temporal variation over the years and across solar cycles. Section \ref{sec:icme_plasma_cool_hot} compares the plasma characteristics of heating and Cooling MEs over solar cycles. Section \ref{sec:dsea} introduces the application of the Superposed Epoch Analysis (SEA) technique to a large set of ICMEs to derive median plasma parameters, not only for the ICMEs but also for the surrounding ambient solar wind, to more effectively characterize ICME properties. Furthermore, we present SEA results for various subsets of ICMEs categorized based on parameters such as the presence of an MC, thermal state, and GS strength. Finally, in Section \ref{sec:discussion}, we summarize the key findings of this study.

\begin{figure*}
    \centering
    \includegraphics[width=0.9\linewidth]{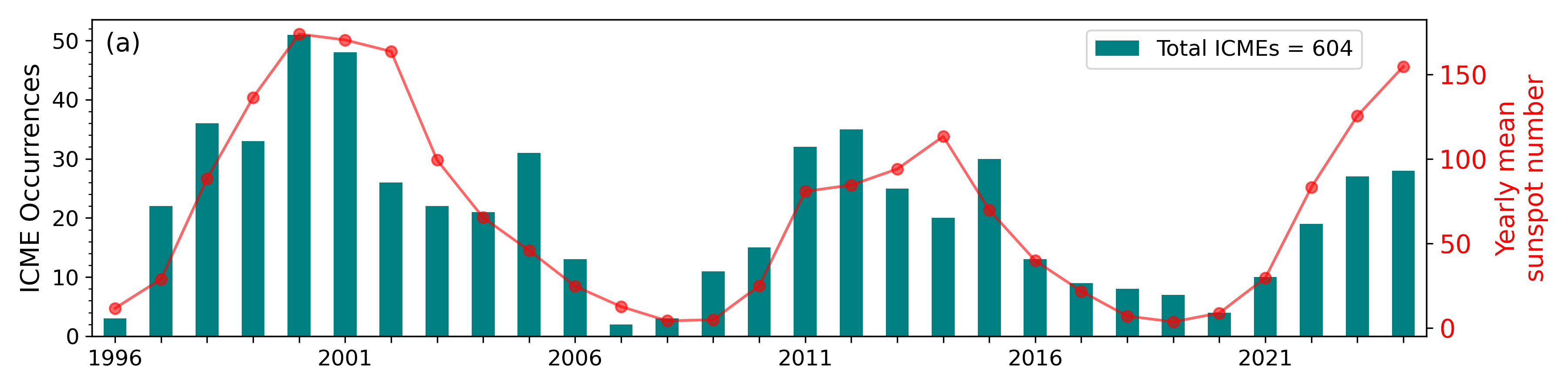}\\
    \vspace{-8pt} 
    \includegraphics[width=0.9\linewidth]{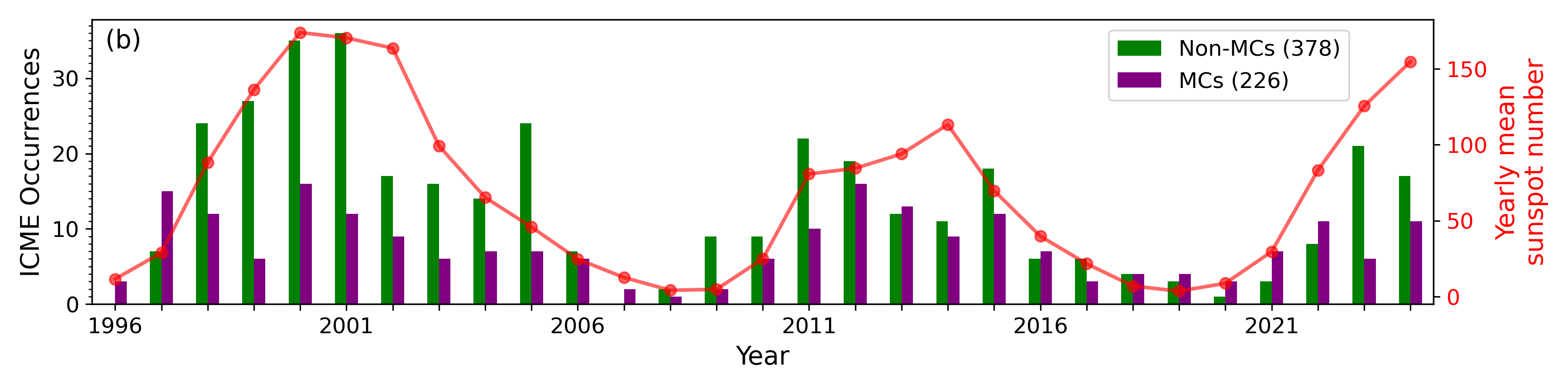}
    \caption{Annual occurrences of (a) ICMEs and (b) MC and non-MC events across Solar Cycles 23, 24, and the rising phase of 25. Yearly mean sunspot numbers are overplotted in both panels for comparison with solar activity levels.}
    \label{fig:icme_no}
\end{figure*}

\section{Data} \label{sec:data}

We used the OMNI database \citep{King2005}, available through NASA’s Space Physics Data Facility (SPDF) at Goddard Space Flight Center (\url{http://omniweb.gsfc.nasa.gov/}). OMNI offers multi-resolution plasma and magnetic field data from 1963 to the present, combining observations from multiple spacecraft and ground stations. The dataset is time-shifted to 1 AU, enabling consistent interplanetary analysis. Its long time span and consistency make it suitable for studying variations in solar wind plasma and the interplanetary magnetic field (IMF) across multiple solar cycles.

In this study, we utilized the continuously updated RC catalog due to its extensive temporal coverage and inclusion of recent events, making it well-suited for analyzing ICME properties across Solar Cycles 23, 24, and the rising phase of Cycle 25. This catalog compiles ICMEs arriving at 1 AU identified visually since 1996 and provides key timing information, including the onset of associated disturbances (typically marked by sudden storm commencements) as well as the start and end times of ME intervals. Notably, the catalog also categorizes the nature of each ME using a flag system: a value of ‘2’ denotes a well-defined MC, ‘1’ indicates partial MC-like characteristics, and ‘0’ reflects an absence of typical MC features. For the purpose of this study, we treat events flagged as ‘1’ or ‘0’ as non-MC events to maintain a clear distinction between classic MCs and other ejecta structures.

It is important to acknowledge that various authors adopt different combinations of parameters to define ICME start and end times, leading to some degree of subjectivity in exact boundaries selection (see discussions in \citealt{Richardson2010}). In particular, identifying the end of the ME is notably challenging as the expected return to pre-ICME conditions, marked by a gradual increase in temperature and a decrease in magnetic field strength, often occurs through a smooth and extended transition without a sharp discontinuity. This results in an hour of timing uncertainties for boundaries in the RC catalog\citep{Richardson2024}. Therefore, there will be some inherent uncertainty in estimating ICME parameters over their entire durations. However, our focus is not only to find the average properties of individual events but also to find the average properties over many events. It is expected that small uncertainties in defining individual ICME boundaries will average out in the statistical analysis involving multiple ICMEs.

\section{Thermal State of magnetic ejecta at 1 AU over solar cycles 23-25} \label{sec:icme_thermal}

\subsection{Annual Occurrence of ICMEs and Their Types} \label{sec:icme_no}

A brief overview of the occurrence of ICMEs and their types provides helpful context on the distribution of events across solar cycles. Figure \ref{fig:icme_no}a shows the annual number of ICMEs from June 1996 to December 2024, along with the yearly mean sunspot number. A total of 604 ICMEs have been identified at Earth, as recorded in the RC catalog.  As expected, a strong positive correlation (Pearson coefficient = 0.83) is observed between ICME occurrence and solar activity. For instance, only three ICMEs were identified during the solar minimum of 1996, compared to 51 in 2000 and 48 in 2001. This upward trend is mostly consistent, except for a notable dip in 1999. The reduced ICME rate that year, also reported by \citep{Cane2000a}, is attributed to an increased presence of co-rotating high-speed streams from low-latitude coronal holes and a restructuring of the near-ecliptic solar wind \citep{Luhmann2002, Richardson2002}. This shift in solar wind conditions led to a temporary decrease in CME-related activity detected near Earth \citep{Cane2003}. Moreover, \citep{Wu2006} reported that an unusually large heliospheric current sheet (HCS) tilt angle and high-latitude prominence eruptions in 1999 likely reduced the number of Earth-directed CMEs. Compared to Solar Cycle 23 (SC23), the number of ICMEs decreased in SC24. This is also shown in several earlier studies confirming the weaker SC24 than the SC23 in terms of CME occurrence rate, mass loss rate via them, their expansion speeds, and geo-effective parameters at 1 AU \citep{Gopalswamy2015,Mishra2019a,Gopalswamy2020,Mishra2024}. Although the sunspot numbers in Solar Cycle 25 (SC25) are more comparable to SC23 and clearly higher than SC24, the ICME count during the rising phase of SC25 appears more similar to that of SC24.

As discussed in Section~\ref{sec:data}, we classify the ICMEs into two categories: MC and non-MC. Over the study period, we identified a total of 226 MC events and 378 non-MC events. Figure~\ref{fig:icme_no}b shows their annual distribution. While SC23 had more non-MCs, SC24 showed a more balanced distribution. Interestingly, the number of MCs remained relatively constant across SC23 and SC24 and did not strongly correlate with sunspot number (Pearson coefficient = 0.62 for MCs vs. 0.82 for non-MCs). This supports previous results that MCs are more common during solar minimum \citep{Cane2003,Jian2006,Richardson2010,Mishra2021a}, likely due to the interaction between CMEs and other structures, such as HCS, which can alter the characteristics of MCs, and is less common during solar minimum \citet{Wu2015mc}.

\begin{figure*}
    \centering
    \includegraphics[width=1\linewidth]{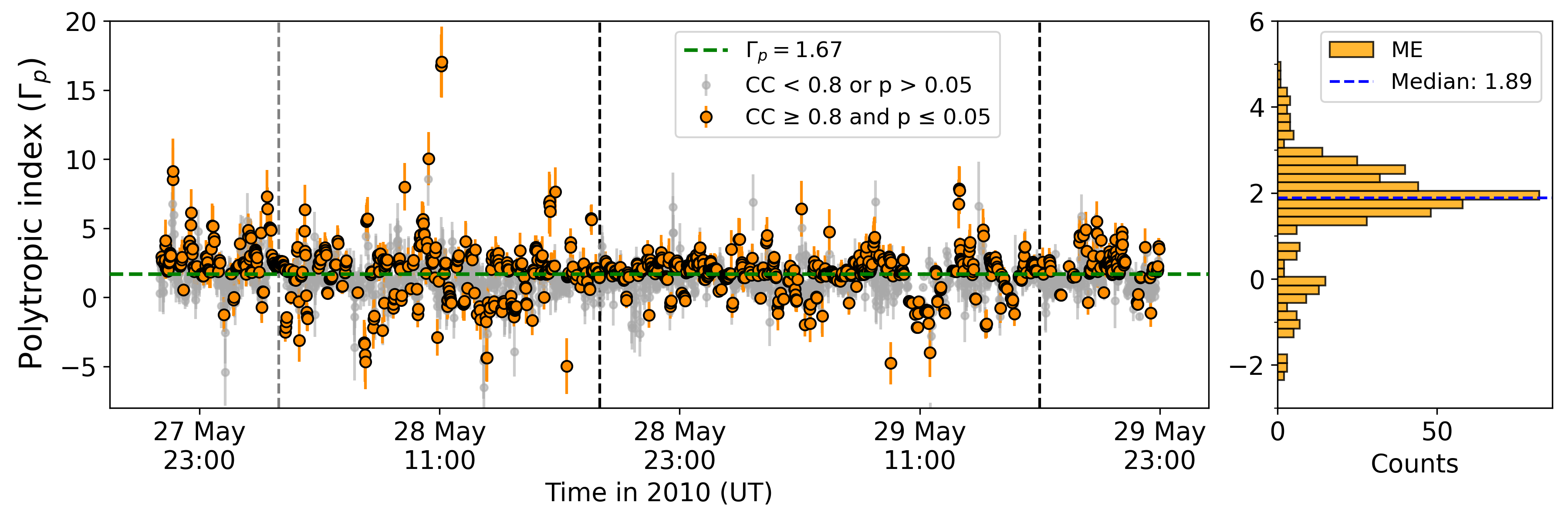}
    \caption{ Derived $\Gamma_p$ values across the ICME event on 28 May 2010. The gray dashed vertical line marks the start of the sheath region, while the black dashed vertical lines denote the boundaries of ME. Orange dots represent reliable $\Gamma_p$ values, whereas gray dots indicate less reliable estimates. The error bars represent the uncertainty due to measurement uncertainty in $N_p$ and $T_p$. The green dashed horizontal line marks the adiabatic index ($\Gamma_p = 5/3$) for reference. The right panel shows the histogram of reliable $\Gamma_p$ values within ME. }
    \label{fig:one_gamma}
\end{figure*}

\subsection{Measurement of Polytropic Index at 1 AU from In-situ Measurements}\label{subsec:poly}

We used 1-minute resolution proton number density ($N_p$) and temperature ($T_p$) data from the OMNI database, accessed via the \textit{ Coordinated Data Analysis Web (CDAWeb): \url {https://cdaweb.gsfc.nasa.gov/}}, to estimate the proton polytropic index ($\Gamma_p$). This index was derived by performing a linear fit between $\log T_p$ and $\log N_p$ and evaluating the fit quality using the Pearson correlation coefficient (CC) and p-value (p). To enhance the reliability of $\Gamma_p$ estimates, we applied this log-log fitting within moving sub-intervals using a 1-minute step size (same as the data resolution). We tested various sub-interval lengths from 3 to 10 data points and found that a 6-point window provided the highest number of statistically robust fits, defined by CC $\geq$ 0.8 and p $\leq$ 0.05. This duration offered an optimal balance, avoiding overly short windows, which can increase statistical noise, and longer ones, which risk including multiple plasma structures, thereby weakening the correlation between $T$ and $n$. This filtering approach ensures that the derived $\Gamma_p$ values reflect the localized thermal state of the plasma within the ejecta and has been used in recent studies \citep{Khuntia2025a, Khuntia2025b}.

Figure \ref{fig:one_gamma} shows the resulting $\Gamma_p$ profile for an ICME, as a representative case, observed on 28 May 2010. The ME is marked by black dashed vertical lines, while the sheath region begins at the gray dashed line. For context, the plot also includes 6 hours of pre and post-ICME ambient solar wind. A value of $\Gamma_p = 5/3$ corresponds to adiabatic expansion with no heat exchange. $\Gamma_p > 5/3$ indicates a heat-release state, whereas $\Gamma_p < 5/3$ suggests a heating state, implying the presence of additional heating processes during plasma evolution. For this particular event, the median $\Gamma_p$ value within the ME region was found to be 1.89, indicating dominant localized heat-release states. Similarly, we determined the median $\Gamma_p$ value for each ME listed in the RC catalog over the entire study period (Table \ref{tab:ICME_list} in Appendix \ref{App_sec:list}). It is important to note that in skewed distributions, the median offers a more reliable measure of central tendency than the mean, as it is less affected by extreme outlier values. Given that the $\Gamma_p$ distributions in each event are generally skewed in our study, the median value more accurately captures the typical thermal behavior of MEs.

Furthermore, since the estimation of $\Gamma_p$ is based on the slope of the linear relationship between $\log T_p$ and $\log N_p$, any measurement uncertainty in $T_p$ or $N_p$ propagates directly into the derived $\Gamma_p$. For 1-minute OMNI Level-2 data, these fractional uncertainties vary with time and instrument calibration, typically within $\delta N_p = 2$–3 \% and $\delta T_p = 2$–8 \% \citep{King2005, Kasper2006}. To quantify their impact on the derived $\Gamma_p$, we performed a Monte Carlo perturbation analysis by adding Gaussian noise with $\delta N_p = 2 \%$ and $\delta T_p = 5 \%$ to the measured quantities. For each 6-point window, $\Gamma_p$ was recalculated 1000 times, and the standard deviation of the resulting distribution was taken as the 1-$\sigma$ uncertainty for that window. The median of all window-level uncertainties within ME was adopted as the representative event-level uncertainty, reported in the seventh column of Table \ref{tab:ICME_list}. Considering all 598 events, the overall median uncertainty in $\Gamma_p$ is found to be 0.41.  An example of this propagation is shown for the 28 May 2010 ICME in Figure \ref{fig:one_gamma}, where the window-wise uncertainties are plotted as error bars, yielding a median uncertainty of 0.45 for the event. As expected, larger measurement errors in $T_p$ and $N_p$ result in correspondingly higher uncertainty in $\Gamma_p$ \citep{Nicolaou2019, Nicolaou2020}.

Some fraction of windows exhibit negative values of $\Gamma_p$, as seen in Figure~\ref{fig:one_gamma}. Physically, such values do not correspond to a classical thermodynamic state but rather indicate an inverse relationship between plasma pressure and density, where the pressure (or temperature) increases as the density decreases. This behavior can arise when non-adiabatic or magnetically dominated processes govern the local plasma dynamics within an ICME. In particular, negative $\Gamma_p$ values may occur in regions where magnetic tension or flux-rope expansion redistributes plasma in a way that the thermal pressure locally decreases despite compression, or where continuous energy injection (e.g., from turbulence, wave–particle interactions, or reconnection-driven heating) preferentially enhances the temperature in lower-density regions. Hence, negative $\Gamma_p$ values are interpreted not as physically negative thermodynamic indices but as indicators of strongly non-equilibrium, magnetically controlled, or anisotropic energy exchange within the CME–solar wind system. Their occurrence marks intervals where classical adiabatic or isothermal assumptions fail and where the coupling between magnetic tension, expansion, and thermal energy exchange is most dynamic.

%%%%%%%%%%%%%%%%%%%%%%%%%%%%%%%%%%%%%%%%%%%%%%%%%%%%%%%%%%%%%%%%%%%%%
\subsection{Yearly Variation of ME's Thermal State at 1 AU}\label{subsec:yearly_poly}

\begin{figure*}
    \centering
    \includegraphics[width=1\linewidth]{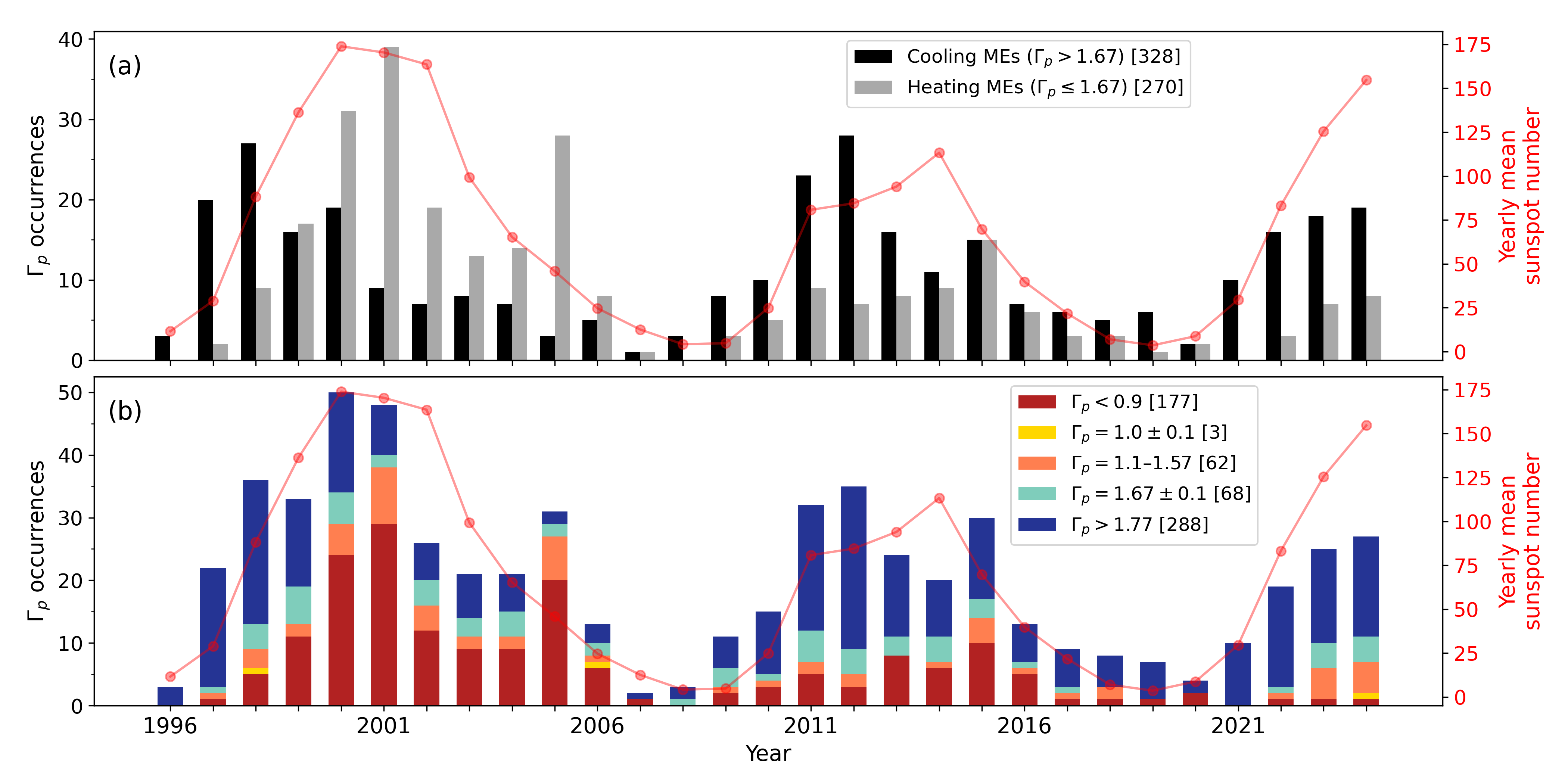}
    \caption{(a) Annual occurrence of heating and cooling ME across SC23, 24 and rising phase of SC25. (b) Annual distribution of ME with various categories depending on $\Gamma_p$ values.}
    \label{fig:yearly_gamma}
\end{figure*}

To examine the yearly variation in the dominant thermal states of MEs across Solar Cycles, we classified the events into two categories: (i) Heating MEs (with median $\Gamma_p \leq 1.67$) and (ii) Cooling MEs (with median $\Gamma_p > 1.67$). Figure \ref{fig:yearly_gamma}a shows the distribution of yearly occurrences of Heating and Cooling MEs. In SC23, during the solar maximum and decline phase, the majority of MEs exhibited a heating state. However, during the rising phase of SC23, the thermal state shifted, with a dominance of heat-release states for MEs. A similar pattern continues into the rising phase of SC24 and SC25, where a major fraction of MEs exhibits a heat-release state. Thus, it is noticeable that during the rising phases, MEs predominantly exhibit a heat-release state, whereas during the declining phases, MEs show a more balanced state or lean towards a heating state. This thermal evolution trend aligns with previous statistical studies on ME composition on SC23, which reported that elemental charge states and relative abundances tend to increase during the rising phase and remain elevated throughout solar maxima and descending phases, compared to the lower values observed during solar minima \citep{Gu2020, Song2021}. The ionic charge states of plasma within a CME serve as a diagnostic imprint of the electron temperature in the source region \citep{Lepri2013}. Since heavy ion charge states and elemental compositions are largely preserved during CME propagation \citep{ko2010, Gruesbeck2011, Lepri2012}, this similarity in the trends of composition and thermal state suggests that CMEs occurring during solar maxima and descending phases originate with inherently higher electron (and possibly proton) temperatures. As these CMEs propagate to 1 AU, the elevated thermal conditions at their origin likely contribute to the observed heating states in protons, indicating a strong linkage between the CME source environment and their thermal behavior in the interplanetary medium.

The detailed classification of MEs into distinct thermal states based on the polytropic index $\Gamma_p$, Major-heating ($\Gamma_p < 0.9$), Isothermal ($\Gamma_p = 1.0 \pm 0.1$), Heating ($1.1 \leq \Gamma_p \leq 1.57$), Adiabatic ($\Gamma_p = 1.67 \pm 0.1$), and Major-cooling ($\Gamma_p > 1.77$), provides more nuanced insights into their thermodynamic evolution across Solar Cycles (Figure \ref{fig:yearly_gamma}b). These categories represent varying thermal behaviors: Major-heating implies active heating where the temperature of the ME can increase during expansion; Isothermal reflects conditions where the ME maintains an almost constant temperature; Heating indicates a slower temperature decrease compared to adiabatic cooling, suggestive of less energy input compare to expansion heat loss; Adiabatic represents an ideal expansion with no heat exchange, leading to a temperature decrease due to expansion-related work done; and Major-cooling denotes enhanced heat-loss processes where the temperature drops more rapidly than adiabatically.

In SC23, Major-heating events show a clear rise during the ascending phase, from 1 in 1997 to 5 in 1998 and 11 in 1999 (Table \ref{tab:thermal_state_perce}). Yet, this phase is dominated by Major-cooling events, particularly in 1997–1998, comprising 86\% and 64\% of total events, respectively. A shift occurs during 2000–2001, when Major-heating events account for ~48–60\% of all cases (Figure \ref{fig:yearly_gamma}b), reflecting strong CME-associated heating during peak activity. Post-maximum, counts decline sharply (12 in 2002, 9 in 2003), but 2005 shows a resurgence to 20 events (~65\%), indicating prolonged heating activity in the descending phase. Overall, this pattern suggests enhanced energy input into MEs during SC23, likely linked to elevated source-region plasma temperatures near solar maximum and descending phase.

In contrast, SC24’s ascending phase (2008–2012) is dominated by Major-cooling events, peaking at 74\% in 2012, implying stronger heat loss or reduced heating efficiency. The 2013–2014 maximum exhibits a balanced mix of heating and cooling, unlike SC23’s heating-dominated peak, which highlights the distinct thermal behavior during the solar maxima of SC23 and SC24. This distinction could stem from the inherent properties of CMEs during their eruption or from variations in the interplanetary medium through which the CMEs propagate, which may influence their evolution and affect their thermal state at 1 AU. The CMEs' expansion speeds at 1 AU and the properties of the interplanetary medium during the maximum of SC23 and SC24 are found to be different in earlier studies \citep{Gopalswamy2015, Lamy2019, Gopalswamy2020, Mishra2021a}. In the early rising phase of SC25 (2020–2024), Major-cooling events remain prevalent (86\% in 2019, 84\% in 2022), though 2023–2024 show a modest rise (24–27\%) in heating-related states, hinting at renewed thermal activity as the cycle strengthens. Notably, isothermal MEs are rare across all years, appearing sporadically and constituting a negligible fraction of the total, reinforcing that MEs at 1 AU are typically thermodynamically active and deviate significantly from equilibrium expansion.

These results emphasize that the thermal states of MEs are not uniformly distributed across the solar cycle but are closely modulated by the phase of solar activity. The dominance of Major-heating states during solar maxima and descending phases suggests stronger heating at the CME source regions, likely from flare-associated energy release or shock-driven compression, which is partly retained during heliospheric propagation. Conversely, the prevalence of Major-cooling states during rising phases may reflect the emergence of CMEs with cooler coronal origins or insufficient internal energy to offset adiabatic expansion and radiative losses. Thus, this classification not only captures the thermal behavior of MEs but also serves as a diagnostic of the large-scale thermodynamic evolution of the solar wind-CME system over solar cycles. 

\begin{figure*}
    \centering
    \includegraphics[width=1\linewidth]{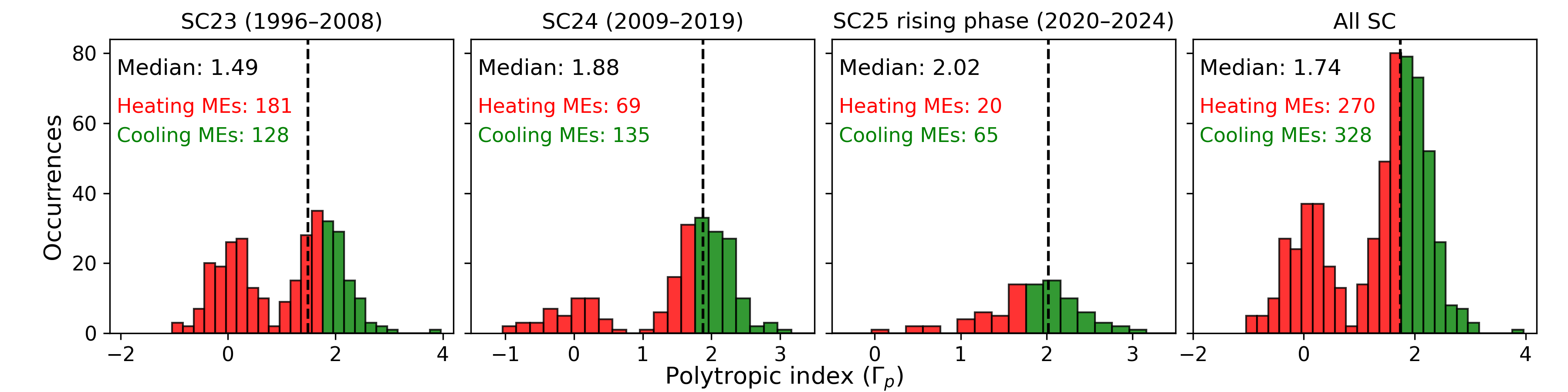}
    \caption{Distribution of $\Gamma_p$ over in SC23, 24 and ascending phase of SC25}
    \label{fig:sc_gamma}
\end{figure*}

Overall, our findings suggest the coexistence of heating and cooling MEs reflects the complex thermodynamic interaction between an expanding flux rope and the surrounding solar wind. Heat exchange in the ME–solar wind system can primarily occur through a combination of adiabatic expansion, thermal conduction along magnetic field lines, and turbulent or compressive energy dissipation. During propagation, a freely expanding ME tends to cool adiabatically as its internal pressure and density decrease, corresponding to $\Gamma_p$ values approaching the adiabatic limit ($\sim$5/3). However, in many cases, the derived $\Gamma_p$ is significantly lower, implying net heat input into the ME plasma. The physical mechanisms responsible for such heating are not yet fully understood, but several plausible processes have been proposed. One possibility is the transfer of heat from the lower corona to the CME flux rope through its magnetic footpoints that remain connected to the Sun \citep{Larson1997}. Continuous energy injection from the ambient solar wind into the CME in the outer corona may also contribute, although cross-field diffusion of charged particles is expected to be strongly inhibited by the magnetic field geometry \citep{Zhang2003}. Magnetic energy dissipation at different heights within the CME can also contribute to heating, potentially influencing its global kinematics and expansion rate. The untwisting of the flux rope may release magnetic energy that sustains CME expansion \citep{Vourlidas2000}, while a portion of the internal magnetic energy could be converted into heat via Joule dissipation ($j^2/\sigma$), where $j$ is the current density and $\sigma$ is the electrical conductivity \citep{Kumar1996}. Although the high conductivity of the interplanetary medium limits the efficiency of Joule heating, internal magnetic reconnection within the CME \citep{Farrugia1993} or between the CME and the interplanetary magnetic field \citep{Lugaz2013} may still play a significant role in converting magnetic energy into thermal energy. These processes, together with conductive and turbulent energy transfer from the surrounding solar wind \citep{Akmal2001, Ciaravella2001, Bemporad2007, lee2009, Landi2010, Manchester2017, Temmer2021}, collectively can determine whether an ME undergoes net heating or cooling during its interplanetary evolution. This thermodynamic diversity underscores the importance of ME–solar wind coupling in shaping CME propagation. Therefore, future studies should focus on quantifying the heat-generation efficiency of these different mechanisms and assessing their relative contributions to the thermal evolution of CMEs throughout the heliosphere.

\subsection{Solar Cycle Variation of ME's Thermal State at 1 AU}\label{subsec:sc_poly}

Figure \ref{fig:sc_gamma} presents SC variation in the distribution of median $\Gamma_p$ values for each ME. In SC23, the median $\Gamma_p$ value is found to be 1.49, indicating a predominance of Heating MEs. Specifically, 181 out of 309 MEs (approximately 59\%) exhibited a heating state at 1 AU. In contrast, SC24 shows a median $\Gamma_p$ of 1.88, suggesting a shift toward MEs with a heat-release (cooling-dominated) state. Only 69 out of 204 MEs (about 34\%) showed heating characteristics during this cycle. In the rising phase of SC25, the median $\Gamma_p$ further increases to 2.02, indicating an even stronger prevalence of heat-release states. Here, only 20 out of 85 MEs (around 23\%) exhibited heating signatures.

These results suggest a significant solar cycle dependence in the thermal state of ICMEs, with heating being more dominant during SC23 and gradually declining in SC24. The overall median $\Gamma_p$ across all three solar cycle phases is 1.74, which is slightly above the adiabatic index of $5/3$ (1.67). This proximity to the adiabatic limit indicates that, on average, the plasma within MEs at 1 AU is neither strongly heated nor strongly cooled, although notable variations exist. Notably, 270 out of 598 MEs (approximately 45\%) still exhibit heating behavior, underscoring the complexity and variability of CME thermodynamics across different heliospheric conditions. Interestingly, a recent statistical study by \citet{Katsavrias2025} reports a similar average polytropic index of $\Gamma = 1.63$ for MEs. Their analysis uses a different CME catalog \citep{Teresa2019}, spanning 1995 to 2001, along with Wind spacecraft data, and applies a slightly different set of filters to estimate $\Gamma$. It is worth noting that their value represents an average, whereas our study focuses on the median of the $\Gamma$ distribution. In our study, we notice a clear bimodal distribution in $\Gamma_p$, with a noticeable dip around $\Gamma_p = 1$ (Figure~\ref{fig:sc_gamma}), consistent with the findings of \citet{Katsavrias2025}. However, isolating the true isothermal state is challenging due to filtering effects that can distort the underlying distribution. Such filtering can artificially suppress tails or create enhanced dips near $\Gamma_p = 1$, which may be further exaggerated by parametric fitting approaches (e.g., Gaussian or $\kappa$-Gaussian), as these impose rigid functional forms that may not reflect the true complexity of the ICME data. To mitigate these issues, we adopt a non-parametric, assumption-free approach by using the median and its bootstrap-derived uncertainty to characterize the central tendency of $\Gamma_p$. Unlike parametric fitted estimates, the median is robust to skewness and local depletions, as it relies only on rank ordering rather than density within narrow intervals. This ensures that, unless filtering drastically affects values near the central rank (which is not the case in our data), the median remains a stable and unbiased measure. Overall, this methodology allows for a more reliable interpretation of thermodynamic variability across ICMEs, particularly in the presence of filtering artifacts and deviations from unimodality.

We find that CME thermodynamics are modulated by solar cycle activity, with variations likely arising from differences in CME initiation mechanisms, coronal background conditions, and CME–solar wind interactions. However, it remains challenging to determine which of these factors primarily governs the thermal state of MEs observed at 1 AU. Previous studies have shown that a higher fraction of CMEs in SC24 originated from non-active region (non-AR) sources, such as filament eruptions from quiet Sun regions and higher latitudes, compared to SC23 \citep{Gopalswamy2016, Lamy2019,Mishra2019a}. Additionally, the expansion behavior of ICMEs differed significantly between the two cycles due to a reduced heliospheric background pressure in SC24 \citep{Gopalswamy2020, Mishra2021a}. This enhanced expansion likely contributed to the observed cooling-dominated thermal state during SC24. There is also a possibility that the MEs were inherently different at the time of the eruption, depending on their source regions or associated flare characteristics \citep{Oloketuyi2019, Lin2023, Pandey2023}. Furthermore, the frequency and intensity of GS were markedly lower in SC24 compared to SC23, which may be a reflection of less energetic CME events during that cycle \citep{Mishra2024}. These factors together offer a plausible explanation for the transition from heating to cooling-dominated MEs from SC23 to SC24.

\begin{figure*}
    \centering
    \includegraphics[width=0.95\linewidth]{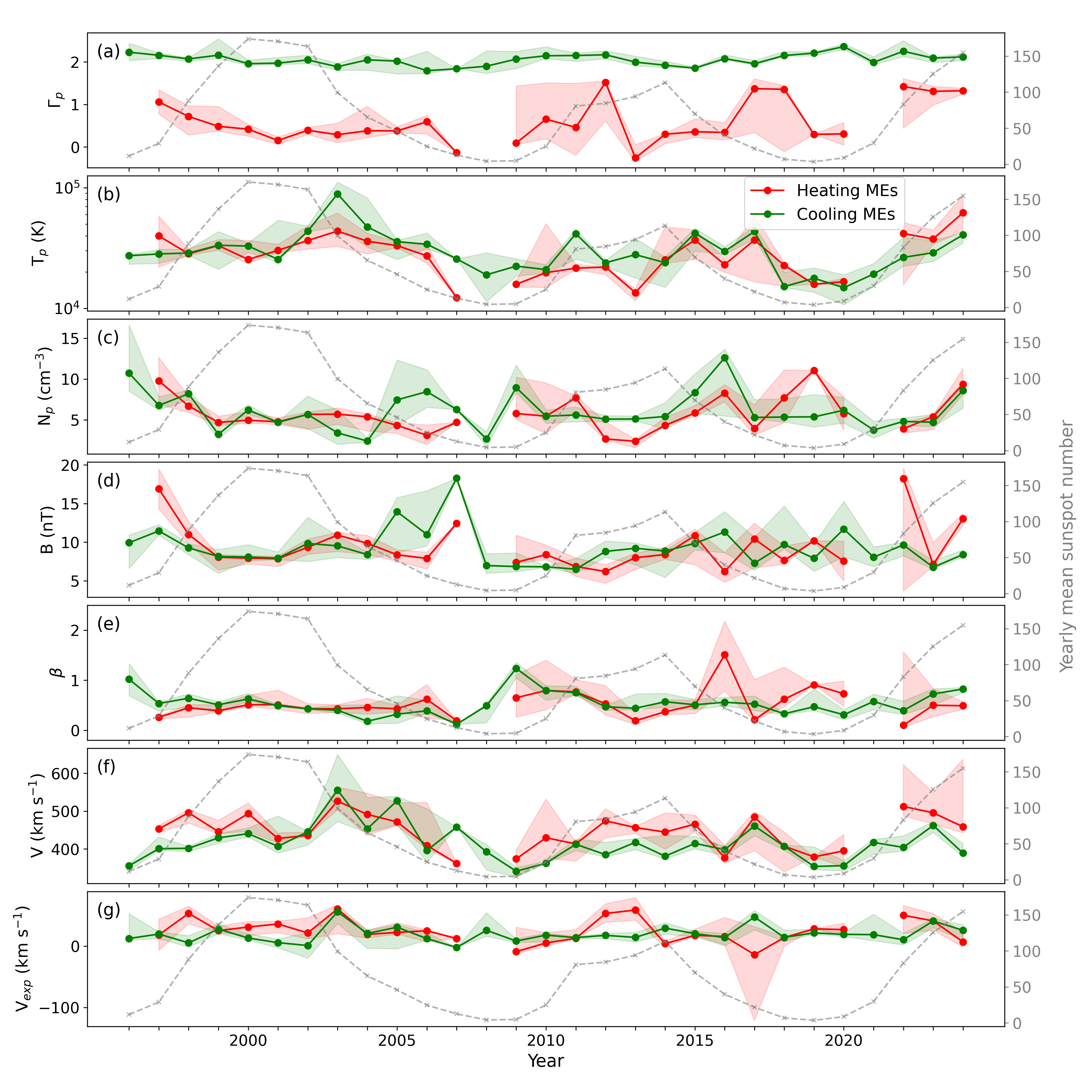}
    \caption{Yearly median values of (a) polytropic index ($\Gamma_p$), (b) proton temperature ($T_p$), (c) proton number density ($N_p$), (d) magnetic field strength ($B$), (e) plasma beta ($\beta$), (f) bulk speed ($V$), and (g) expansion speed ($V_{exp}$) for heating and Cooling MEs across Solar Cycles 23, 24, and the ascending phase of Cycle 25. The background color-shaded regions represent the 68\% (1$\sigma$) confidence intervals obtained via bootstrap resampling. Each panel is overplotted with the yearly mean sunspot numbers to investigate the correlation between solar activity and the thermal, plasma, and magnetic properties of MEs.}
    \label{fig:yearly_heating_cooling_plasma}
\end{figure*}

\section{Variations in plasma properties of heating and cooling magnetic ejecta} \label{sec:icme_plasma_cool_hot}

We classified MEs into two categories based on their $\Gamma_p$ values: Heating MEs ($\Gamma_p < 1.67$) and Cooling MEs ($\Gamma_p > 1.67$). For each category, the yearly median $\Gamma_p$ was computed (Figure~\ref{fig:yearly_heating_cooling_plasma}a). We found that Cooling MEs consistently exhibit a median $\Gamma_p \approx 2$ with minimal year-to-year variation, whereas Heating MEs show substantial variability but no clear correlation with solar activity (represented by annual sunspot numbers). During SC23, $\Gamma_p$ for Heating MEs decreases from the ascending phase to solar maximum and remains nearly constant during the descending phase, consistent with the enhanced occurrence of Heating MEs in those periods (Figure~\ref{fig:yearly_gamma}). In SC24, $\Gamma_p$ for Heating MEs initially rises in the ascending phase, drops sharply around the 2013 maximum, and then gradually increases toward the 2018 minimum. In the rising phase of SC25, $\Gamma_p$ remains nearly constant, close to an isothermal state. Overall, Heating MEs exhibit stronger heating (lower $\Gamma_p$) during solar maxima and descending phases, while during solar minima (e.g., 2008 and 2021), only Cooling MEs are observed with $\Gamma_p \approx 2$, indicating dominant heat loss.

Figure~\ref{fig:yearly_heating_cooling_plasma}b shows the yearly median $T_p$ values for both ME categories. For Cooling MEs, $T_p$ remains nearly constant through the ascending and maximum phases of SC23, rises slightly during the early descending phase, and then declines toward the 2008 minimum. A similar but more irregular trend appears in SC24, with $T_p$ increasing through the ascending and maximum phases and decreasing thereafter. Starting from the deep minimum in 2008, $T_p$ rises gradually until 2017, falls in 2018, and then increases again through the rising phase of SC25. This long-term variation reflects solar cycle modulation, higher temperatures during active phases, and lower values during minima, consistent with earlier findings \citep{Chi2016, Jian2018}, which reported higher solar wind temperatures during active phases and lower values during minima.

Heating MEs display a broadly similar trend in $T_p$ as Cooling MEs, with a rising pattern during the ascending, maximum, and early descending phases, followed by a declining phase. However, certain interesting differences are evident. Notably, in 2013, Figure \ref{fig:yearly_heating_cooling_plasma}a shows a significantly lower $\Gamma_p$ value for Heating MEs, indicating stronger thermal heating during propagation. Yet, the in-situ measured $T_p$ for the same year is lower compared to neighboring years. This observation highlights an important physical insight: the in-situ measured $T_p$ is a combined result of the initial eruption temperature of the ME and the subsequent heating it experiences during its interplanetary journey. Thus, despite stronger in-transit or ongoing heating (as suggested by the low $\Gamma_p$), the lower $T_p$ implies that the MEs erupted with a relatively cooler initial temperature in 2013. A similar behavior is noticed during solar minima years like 2007-2009 and 2019–2020, where $\Gamma_p$ values are lower (indicating more heating), but $T_p$ is also relatively low. This suggests that during solar minima, MEs generally erupt with lower inherent temperatures, likely due to weaker solar activity with lower CME energetics \citep{Vourlidas2011err,Gopalswamy2014,Lamy2019,Pant2021}. Thus, while heating processes during propagation contribute significantly to the thermal state of MEs, the initial temperature at eruption, strongly modulated by solar activity levels, plays a crucial role in determining the final in-situ observed $T_p$. Furthermore, when we compared the annual median $T_p$ between heating and Cooling MEs during SC23 and SC24, we found that the Cooling MEs generally exhibit higher $T_p$ values than the Heating MEs. In contrast, during the ascending phase of SC25, the Heating MEs show higher $T_p$ compared to the Cooling MEs. This change may reflect a shift in the source region properties or heliospheric conditions between SCs.

Figure~\ref{fig:yearly_heating_cooling_plasma}c presents the yearly median values of $N_p$ for Heating MEs and Cooling MEs. During SC23, Cooling MEs exhibit a generally declining and somewhat irregular trend in $N_p$ throughout the ascending, maximum, and early descending phases, continuing until 2004. From 2005 to 2007, $N_p$ displays an enhanced level, followed by a notable decrease during the deep solar minimum in 2008. In SC24, $N_p$ for Cooling MEs shows a clear rising trend from 2008 to 2016, reaching a peak around 14~cm$^{-3}$. After this peak, $N_p$ maintains a relatively stable level until 2020. A slight decrease is observed afterward, followed by a renewed increase during the ascending phase of SC25. In contrast, Heating MEs exhibit a comparatively smoother evolution in $N_p$ during SC23, with a gradual decline observed up to 2006. Entering SC24, $N_p$ values appear elevated compared to SC23 during the ascending phase but reach a minimum around the solar maximum in 2014. Following this dip, $N_p$ steadily increases up to 2019, with a minor drop occurring in 2017. After 2019, $N_p$ values again decreased slightly, but from 2022 onward, a renewed increasing trend is evident during the rising phase of SC25. Overall, Heating MEs have consistently lower values of  $N_p$ compared to Cooling MEs.

Figure~\ref{fig:yearly_heating_cooling_plasma}d shows the yearly median values of $B$ for Heating and Cooling MEs. For Cooling MEs, $B$ declines during SC23’s ascending and maximum phases, then rises irregularly through the descending phase, peaking near $\sim$18~nT around the 2007 minimum. In SC24, for Cooling MEs, $B$ increases smoothly from 2008 to a peak of $\sim$11~nT in 2016, then decreases toward SC25. Heating MEs exhibit more distinct cycle-dependent variations: $B$ falls sharply during SC23’s early ascent, stabilizes near the maximum, increases slightly during 2001–2003, and rises sharply again near the 2007 minimum. In SC24, $B$ increases from ascent to early descent (peaking around 2015) and remains elevated compared to earlier cycles. In SC25, Heating MEs exhibit elevated and fluctuating $B$ values that are generally higher than those seen in SC24. Overall, the long-term variation of $B$ does not exhibit a clear and systematic solar cycle dependence for either category. However, it is noteworthy that both categories tend to exhibit their maximum $B$ values during the descending phases of the solar cycles. Furthermore, the $B$ profiles of Heating and Cooling MEs are broadly similar in magnitude and variability across the solar cycles, suggesting common underlying physical drivers modulating the magnetic field strength of MEs over time.

We observe a reversal trend in $\beta$ for heating and Cooling MEs during SC23 (Figure~\ref{fig:yearly_heating_cooling_plasma}e). Specifically, $\beta$ decreases gradually for Cooling MEs while it increases steadily for Heating MEs from the ascending to the descending phase of SC23. During the solar minimum around 2007–2009, $\beta$ for Cooling MEs shows an increasing trend. However, during SC24, $\beta$ again decreases from the ascending to the descending phase for Cooling MEs. Unlike the behavior before SC24, we do not observe a significant rise in $\beta$ during the solar minimum between SC24 and SC25. In SC24, during the solar maximum, Cooling MEs exhibit a higher $\beta$ compared to Heating MEs. However, during the descending phase of SC24, the Heating MEs show a higher $\beta$ than the Cooling MEs.

Across all SCs, we found that $V$ increases gradually during the ascending phase and solar maximum, peaking around the early descending phase before decreasing towards the solar minimum for both Heating and Cooling MEs (Figure~\ref{fig:yearly_heating_cooling_plasma}f). Notably, during the ascending and maximum phases, Heating MEs generally exhibit higher flow speeds compared to Cooling MEs. This suggests that Heating MEs during these phases are associated with more energetic eruptions. Moreover, $V_{exp}$ of MEs also shows interesting behavior across SCs (Figure~\ref{fig:yearly_heating_cooling_plasma}g). During solar maxima, Heating MEs tend to have higher expansion speeds compared to Cooling MEs. In particular, for the year 2013, we observe a notable case: despite a lower $T_p$ and lower $\Gamma_p$ (indicating enhanced heating), $V_{exp}$ is higher. This suggests that even though the internal thermal energy of the ME was relatively low (low $T_p$), the outward expansion was significant, likely supported by a strong internal magnetic pressure or a higher eruption energy at the source region.

\section{Superposed Epoch Analysis applied to ICME} \label{sec:dsea}

We employed the Superposed Epoch Analysis (SEA) technique to statistically determine the average temporal evolution of plasma parameters and geomagnetic indices associated with different categories of ME events. SEA aligns multiple events to a common reference time, typically the event start time, and normalizes their durations, allowing characteristic features to emerge by averaging (or taking the median) across events. This approach highlights common trends but may mask individual variations. We use the SEA as a statistical method to determine averaged profiles of physical parameters for the normalized duration of the events. SEA is a widely used technique in space physics and has been applied to study the solar wind and ICME \citep{Zhang1988, Winslow2015sea, MasiasMeza2016, Janvier2019, Regnault2020, Guo2021}.

To perform SEA, we first normalized the time series of each event (i.e., ME) so that the sheath start time (or the disturbance time as in the RC catalog) and the start and end times of ME are mapped onto a common, standardized timescale. This normalization ensures that different events, which may naturally have different durations, can be meaningfully compared on the same timeline. Since MEs size is found to be about three times that of the sheath at 1 AU \citep{Zhang2008,MasiasMeza2016,Mishra2021a}, we assigned different normalized time ranges: we divided the sheath interval into 25-time bins and the ME interval into 75-time bins. This choice preserves the typical size ratio between the sheath and ME regions. In addition to the sheath and ME regions, we also included two more intervals: a pre-ICME solar wind region (before the sheath) and a post-ICME wake region (after the ME). The durations of these intervals were made equal to the sheath and ME durations, respectively. After the normalization, we divided the time series of each ME into an equal number of bins. This is important because each ME event has a different duration and a different number of recorded data points. By binning the data, we ensure that each ME contributes equally to the final profiles. Finally, for each individual time bin across all MEs events, we calculated the median values of the studied parameters, such as magnetic field strength, plasma properties, and dynamic pressure. This method allows us to extract the average behavior of ME structures and their surrounding regions by emphasizing common patterns while minimizing the effect of event-to-event variability.

\begin{figure*}
    \centering
    \includegraphics[width=1\linewidth]{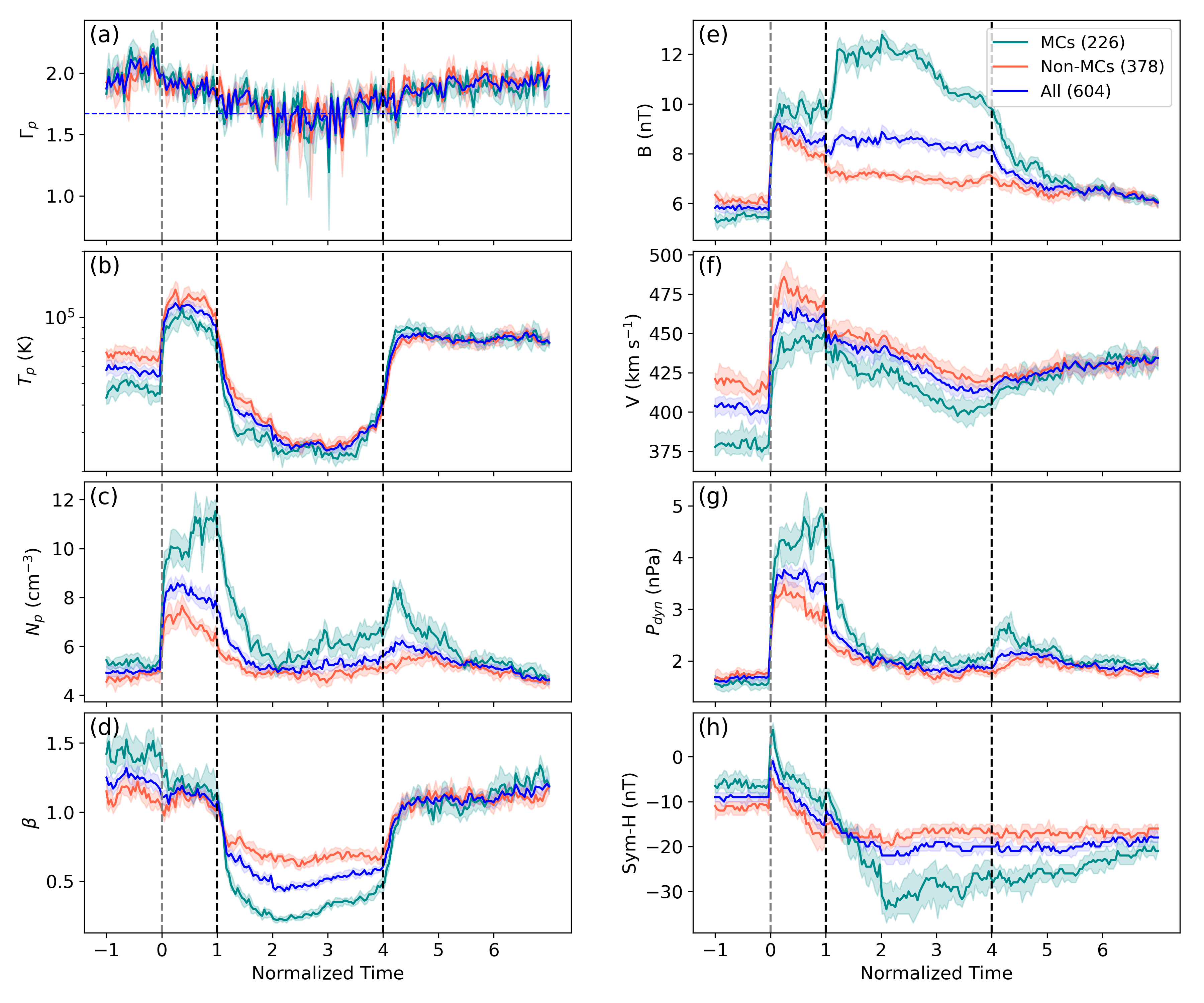}
    \caption{Superposed Epoch Analysis (SEA) showing the median values of parameters across the pre-ICME, sheath, ME, and post-ICME regions: (a) polytropic index ($\Gamma_p$), (b) proton temperature ($T_p$), (c) proton number density ($N_p$), (d) plasma beta ($\beta$), (e) magnetic field strength ($B$), (f) flow speed ($V$), (g) dynamic pressure ($P_{dyn}$), and (h) geomagnetic index (Sym-H). The curves represent: teal for MC, orange for non-MC, and blue for all MEs. Black vertical dashed lines mark the boundaries for the ME region, and gray dashed vertical lines mark the starting sheath region. The background color-shaded regions represent the 68\% (1$\sigma$) confidence intervals obtained via bootstrap resampling.}
    \label{fig:sea_plasma}
\end{figure*}

To thoroughly evaluate differences in the thermodynamic behavior, specifically in the distributions of $\Gamma_p$, across different classes of ICMEs, we also employed a comprehensive set of statistical methods. The \textit{Kolmogorov--Smirnov (KS) test} was used to detect differences in the overall shape and cumulative distribution of $\Gamma_p$; a test statistic $D > 0.1$ combined with a $p$-value less than 0.05 is generally taken as strong evidence that the distributions differ significantly in form. The \textit{Mann--Whitney U test}, a non-parametric test, evaluates whether one distribution tends to produce higher or lower values than another without assuming Gaussian behavior; a $p$-value less than 0.05 similarly indicates a statistically significant difference in central tendency (e.g., medians). To quantify the magnitude and direction of this difference, we used \textit{Cliff’s delta}, which quantifies the effect size by measuring the probability that a randomly selected $\Gamma_p$ value from one group will exceed that from another; values near zero indicate substantial overlap between the two distributions, while values approaching $\pm 1$ suggest nearly complete separation, with one group consistently exhibiting higher or lower values than the other. In addition, we examined the \textit{percentage of time bins} in the SEA where the median $\Gamma_p$ fell below a critical threshold ($\Gamma_p < 1.67$), providing insight into the temporal extent and frequency of enhanced heating. Finally, we applied a \textit{permutation test} on the median $\Gamma_p$ values, which involves randomly reassigning group labels to build a null distribution; if the observed difference lies in the extreme tail (e.g., $p < 0.01$), it strongly suggests the difference is not due to random variation. Together, these complementary tests allow us to assess not only whether two populations differ, but also how strong, localized, and statistically robust those differences are.

\subsection{Evolution in Plasma Properties across MEs (MCs and Non-MCs)}\label{subsec:dsea_plasma}

To investigate the typical behavior of plasma and magnetic field parameters during ICMEs, we performed the SEA using 604 ICMEs from the RC catalog. As discussed before, these events were categorized into magnetic clouds (MCs, flag 2) and non-magnetic clouds (non-MCs, flags 0 and 1), with 226 MCs and 378 non-MCs. Figure \ref{fig:sea_plasma} presents the median profile of parameters such as $\Gamma_p$, $T_p$, $N_p$, $\beta$, $B$, $V$, $P_{dyn}$, and Sym-H in each time bin. The shaded regions in the figure indicate the 68\% (1$\sigma$) confidence intervals for the median values, which were estimated using a non-parametric bootstrap resampling technique (with 1000 iterations). These confidence bands provide a robust, data-driven measure of the statistical variability of the median profiles, accounting for sample dispersion within each time bin. Table \ref{tab:median} provides the median values of these parameters for each region, pre-ICME, sheath, ME, and post-ICME, across different event categories.

We found that $\Gamma_p$ exhibits a distinct four-phase pattern. For all ICMEs, the median $\Gamma_p$ is highest in the pre-ICME region (2.02), dips to 1.83 in the sheath, reaches its lowest in the ME (1.73), and then recovers to 1.91 in the post-ICME region (Table \ref{tab:median}). This behavior indicates a transition from an enhanced cooling state to near-adiabatic expansion in the ME, followed by an enhanced cooling state again. The same trend holds for MCs and non-MCs individually, with MCs having slightly higher $\Gamma_p$ in pre (2.09) and slightly lower in post (1.88) regions compared to non-MCs (1.95 and 1.93, respectively). The comparison between MCs and non-MCs reveals a statistically significant, but very subtle, difference in their $\Gamma_p$ distributions. The KS test yields $D = 0.030$ with a highly significant p-value ($p \approx 1.38 \times 10^{-47}$), indicating a detectable difference in distribution shapes. However, the effect size is small, as supported by Cliff’s Delta of $-0.009$, which implies that a randomly selected non-MC event has a slightly higher $\Gamma_p$ than an MC in only 50.4\% of cases. The Mann–Whitney U test similarly returns a small p-value ($p \approx 2.26 \times 10^{-4}$), reinforcing the statistical significance but not the practical significance.  A permutation test on the median $\Gamma_p$ values further confirms this pattern: the observed median difference between MCs and non-MCs is small ($\Delta \Gamma_p = -0.0519$), yet highly significant ($p < 10^{-5}$). The KDE plot (Figure~\ref{fig:kde}a) confirms this: both distributions show nearly identical bimodal structures across the full range of $\Gamma_p$. The MCs and non-MCs peak similarly around $\Gamma_p \approx 2$, and their overall shapes and tails are closely aligned. These results suggest that although the large sample size detects a statistical difference, the thermodynamic properties inferred from $\Gamma_p$ are largely comparable between MCs and non-MCs. This implies that factors other than magnetic cloud topology may play a more dominant role in shaping the compressive or heating characteristics of ICMEs.

Furthermore, $T_p$ reveals the highest value in the sheath and the lowest values inside the ME (Figure \ref{fig:sea_plasma}b). For all ICMEs, the sheath has the highest median $T_p$ of $11.5 \times 10^4$ K, while the ME has the lowest at $3.0 \times 10^4$ K. This indicates that the compression and turbulence in the sheath significantly increase the plasma temperature, while the ME undergoes substantial cooling due to expansion. When comparing MCs and non-MCs, MCs consistently exhibit lower $T_p$ in pre-ICME ($4.5 \times 10^4 K$  vs. $6.3 \times 10^4 K$), sheath ($10.1 \times 10^4 K$ vs. $12.8 \times 10^4 K$), and ME ($2.8 \times 10^4 K$ vs. $3.1 \times 10^4 K$) regions, indicating that MCs are generally cooler. However, post-ICME $T_p$ are comparable ($7.5  \times 10^4 K$ vs. $7.7 \times 10^4 K$), suggesting similar recovery conditions (Table \ref{tab:median}). $N_p$ shows enhancement in the sheath and a drop in the ME, consistent with expectations from shock compression and subsequent expansion (Figure \ref{fig:sea_plasma}c). The sheath has a median $N_p$ of 8.6 cm$^{-3}$ for all ICMEs, decreasing to 5.3 cm$^{-3}$ in the ME. Interestingly, MCs show even higher $N_p$ in the sheath (11.4 cm$^{-3}$) and ME (6.3 cm$^{-3}$), compared to non-MCs (7.2 and 5.0 cm$^{-3}$ respectively). This implies that MCs are more strongly compressed and carry denser plasma.

The plasma beta ($\beta$), which is the ratio of thermal to magnetic pressure, shows an inverse relationship to $T_p$ (Figure \ref{fig:sea_plasma}d). All ICMEs show a sheath $\beta$ of 1.1, dropping significantly to 0.5 in the ME, indicating magnetic dominance. MCs have an even lower $\beta$ (0.3) than non-MCs (0.7), reinforcing the idea that MCs are more magnetically dominated and this aligns with the magnetic field strength $B$ (Figure \ref{fig:sea_plasma}e). For all ICMEs, the pre-ICME region has a median $B$ of 8.6 nT, comparable to 8.5 nT in the ME. For MCs, the ME region has an even stronger $B$ of 11.3 nT, while non-MCs show a lower 7.1 nT, highlighting the stronger magnetic core in MCs. Notably, $V$ patterns also reveal interesting differences. Non-MCs tend to propagate faster, with ME speeds of 438 km s$^{-1}$ compared to 421 km s$^{-1}$ for MCs. In the sheath, non-MCs are at 479 km s$^{-1}$ vs. 455 km s$^{-1}$ for MCs (Figure \ref{fig:sea_plasma}f). Despite these slower speeds, MCs show higher dynamic pressure ($P_{dyn}$) in both the sheath (4.9 nPa vs. 3.4 nPa) and ME (2.2 vs. 1.9 nPa) (Figure \ref{fig:sea_plasma}g).

Interestingly, the consistently higher $P_{dyn}$ observed in the sheath regions of MCs indicates stronger momentum fluxes associated with these structures. This elevated pressure at the shock arrival often results in a sharp positive spike in the Sym-H index, a phenomenon known as Sudden Storm Commencement (SSC) \citep{Burlaga1969,Veenadhari2012}. Our analysis shows that this positive excursion in Sym-H is generally more prominent for MCs than for non-MCs (Figure \ref{fig:sea_plasma}h). Following the SSC, the Sym-H index typically decreases as the ME passes, particularly when strong southward magnetic fields are present, enhancing magnetic reconnection and intensifying geomagnetic storm activity \citep{Dungey1961,Tsurutani1997,Kamide1997}. While the rate of Sym-H decline in the sheath region is comparable between MCs and non-MCs, a notable difference emerges in the ME region, where MCs exhibit a steeper drop and reach lower minimum Sym-H values. This pattern, initial high $P_{dyn}$ in the sheath triggering SSC, followed by a significant drop in Sym-H during the ME phase, is consistent with the established sequence of geomagnetic storm development. It is evident that MCs tend to produce stronger SSCs and drive more intense geomagnetic storms compared to non-MCs \citep{Wu2007,Echer2008}. Moreover, once the ICME reaches its minimum Sym-H value during the main phase of the storm, the subsequent recovery back to positive values or levels similar to the post-ICME region tends to take longer for MCs than for non-MCs (Figure \ref{fig:sea_plasma}h). This slower recovery suggests a more prolonged disturbance to Earth's magnetosphere following MC-driven storms.

Both MCs and non-MCs exhibit a similar $V_{exp}$ of approximately 20 km s$^{-1}$, indicating that the overall radial growth of the ME structure is nearly identical across these two categories. To examine the internal structure of the ME more closely, we used the distortion parameter (DiP) as defined in \citet{Nieves2018}. Specifically, DiP-B quantifies the asymmetry in the magnetic field strength profile, while DiP-N is an analogous measure for the proton number density profile. A DiP value of 0.5 represents a perfectly symmetric profile centered around the midpoint of the ME. Values less than 0.5 indicate stronger compression at the front portion of the ME, whereas values greater than 0.5 suggest compression toward the rear. For the entire set of ICMEs, we found that both DiP-B and DiP-N have median values of 0.48, implying a slight asymmetry with compression predominantly at the leading edge of the ME. Interestingly, when separating the events into MC and non-MC types, we observed that both have the same DiP-B value of 0.48, suggesting similar magnetic field asymmetry. However, there is a subtle difference in the density structure: MCs show a slightly lower DiP-N of 0.46 compared to 0.49 in non-MCs. This indicates that the density profile within MCs is more front-loaded and asymmetric, whereas non-MCs tend to have a more symmetric density distribution.

\begin{figure*}
    \centering
    \includegraphics[width=1\linewidth]{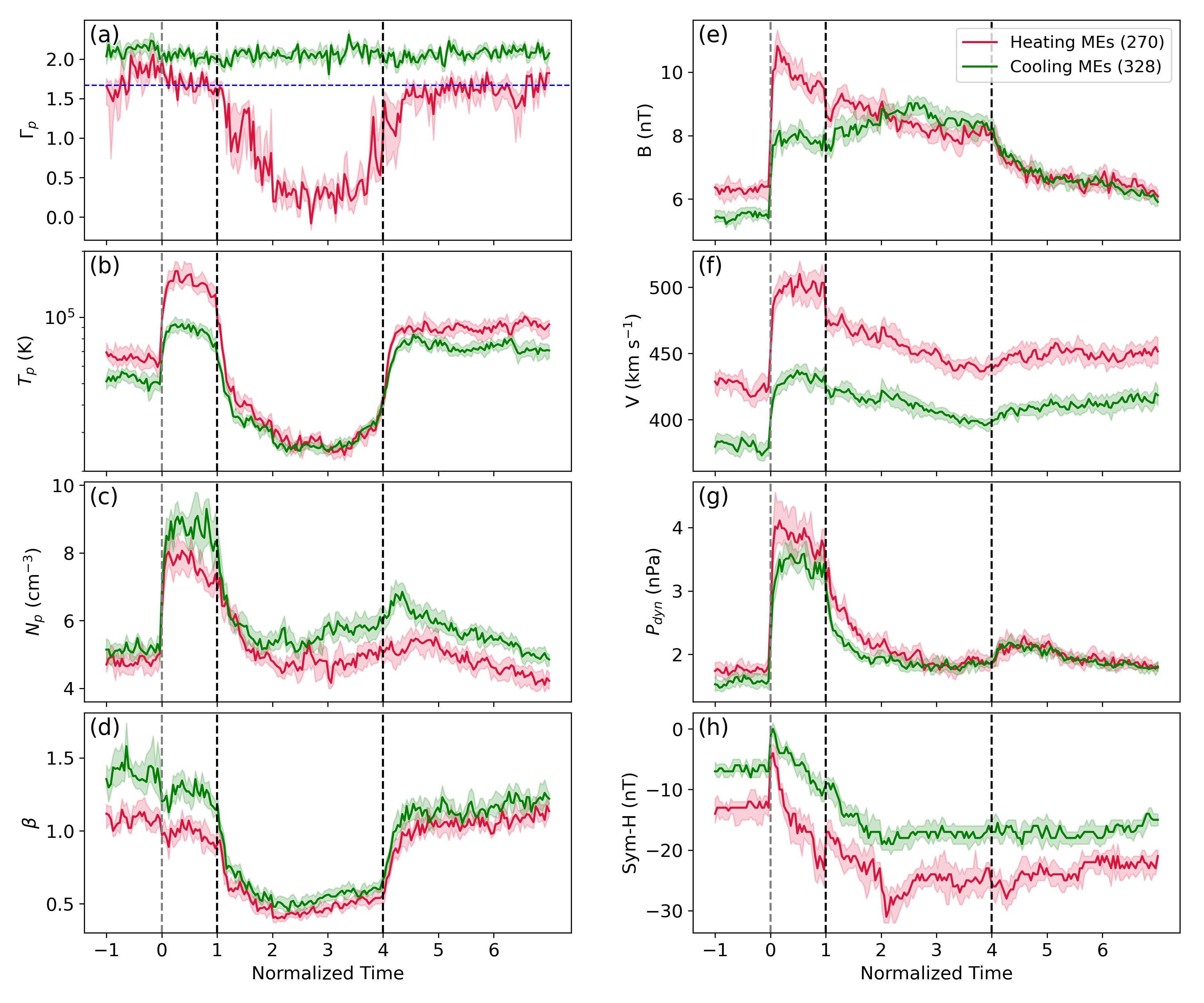}
    \caption{Superposed Epoch Analysis (SEA) showing the median values of parameters across the pre-ICME, sheath, ME, and post-ICME regions: (a) polytropic index ($\Gamma_p$), (b) proton temperature ($T_p$), (c) proton number density ($N_p$), (d) plasma beta ($\beta$), (e) magnetic field strength ($B$), (f) flow speed ($V$), (g) dynamic pressure ($P_{dyn}$), and (h) geomagnetic index (Sym-H). The curves represent: red for Heating ME and green for Cooling ME. Black vertical dashed lines mark the boundaries for the ME region, and gray dashed vertical lines mark the starting sheath region. The background color-shaded regions represent the 68\% (1$\sigma$) confidence intervals obtained via bootstrap resampling.}
    \label{fig:sea_heating_cooling}
\end{figure*}

The SEA reveals systematic plasma and magnetic field variations across ICMEs, with clear differences between MCs and non-MCs. While both categories share broadly similar thermodynamic trends, such as elevated $\Gamma_p$ in the pre- and post-ICME regions, sheath, and near-adiabatic cooling within the ME, MCs are distinguished by cooler temperatures, lower plasma $\beta$, stronger magnetic fields, denser plasma, and higher dynamic pressures, all of which contribute to more pronounced SSCs and deeper Sym-H depressions. Despite these differences, $\Gamma_p$ distributions remain largely comparable between MCs and non-MCs, suggesting that magnetic topology alone does not govern ICME thermal evolution. These results motivate a closer examination of plasma behavior in Heating and Cooling MEs, where thermodynamic states may provide additional insights into their geoeffectiveness.

\subsection{Comparative Plasma Evolution in Heating and Cooling MEs}\label{subsec:dsea_hot_cool}

We performed a comparative SEA between Heating and Cooling MEs based on $\Gamma_p$, where MEs with $\Gamma_p < 1.67$ were categorized as heating and those with $\Gamma_p > 1.67$ as Cooling MEs (Figure \ref{fig:sea_heating_cooling}). This classification reveals important physical differences in plasma and magnetic field behaviors between the two categories across all associated regions, pre-ICME, sheath, ME, and post-ICME, highlighting the underlying thermodynamic and dynamical processes that differentiate the two categories.

For Heating MEs, the evolution of $\Gamma_p$ shows a gradual decline from the pre-ICME region, where it has a median value of 1.84, to 1.64 in the sheath (Table \ref{tab:median}). This decline becomes markedly steep within the MEs, where $\Gamma_p$ reaches a median value of 0.42, indicating significant heating and deviation from adiabatic behavior (Figure \ref{fig:sea_heating_cooling}a). Interestingly, $\Gamma_p$ recovers partially in the post-ICMEs region to 1.59, suggesting the plasma begins to relax after the central heating phase in MEs. In contrast, Cooling MEs exhibit consistently high values of $\Gamma_p$ across all regions, starting from 2.16 in the pre-ICME wind, rising slightly to 2.14 post-ICME, with a minor dip to 2.07 within the ME. This consistently elevated $\Gamma_p$ suggests an enhanced cooling evolution throughout the structure.

The statistical analysis reveals a clear and robust distinction in the distribution of $\Gamma_p$ between Heating and Cooling MEs. The KS test yields a large test statistic ($D = 0.325$) and a p-value effectively equal to zero, indicating that the two distributions differ significantly in shape. The Mann–Whitney U test further supports this with $p < 10^{-10}$, confirming a statistically significant shift in central tendency without assuming Gaussian behavior. Cliff’s Delta of $-0.361$ indicates a medium to large effect size, meaning that a randomly selected Cooling ME has a higher $\Gamma_p$ than a randomly selected Heating ME 68\% of the time. This trend is corroborated by the KDE plot (Figure~\ref{fig:kde}b), where the Cooling population peaks sharply around $\Gamma_p \approx 2$, whereas the Heating population exhibits a broader, bimodal structure with enhanced density at lower $\Gamma_p$ values (near 0), indicative of more compressive or sub-adiabatic behavior. Additionally, a permutation test on the median $\Gamma_p$ values provides additional confirmation: the observed median difference between Heating and Cooling MEs is substantial ($\Delta \Gamma_p = -1.82$), with a p-value effectively zero, reinforcing that this difference is not attributable to random variation. Collectively, these results strongly suggest that Heating MEs are characterized by lower polytropic indices, indicative of enhanced internal energy deposition, while Cooling MEs tend to follow more adiabatic or thermally relaxing trajectories.

We noticed that $T_p$ profile trends reinforce this distinction. Heating MEs show a substantial increase in the sheath region ($15.3 \times 10^4 K$) compared to the pre-ICME solar wind ($6.2 \times 10^4 K$), implying shock or compression-induced heating at the leading edge of the disturbance (Table \ref{tab:median}). While temperature falls to $3.1 \times 10^4 K$ inside the ME, typical for ME interiors, the extremely low $\Gamma_p$ here reveals that this low temperature compared to surrounding regions is not due to adiabatic cooling but possibly due to non-adiabatic energy exchange. The post-ICME region sees a  $T_p$ rebound to $8.6 \times 10^4 K$. On the other hand, Cooling MEs show a smoother $T_p$ evolution: starting at $4.9 \times 10^4 K$ pre-ICME, rising to $9.3 \times 10^4 K$ in the sheath, and dropping to $2.9 \times 10^4 K$ within the MEs, followed by a rise to $6.8 \times 10^4 K$ post-ICME. The cooling is more uniform, matching the stable high values of $\Gamma_p$ ($> 1.67$) across all the regions. Further, $N_p$ illustrates this contrast. In Heating MEs, $N_p$ rises from 5.0 $cm^{-3}$ in pre-ICME to 7.7 $cm^{-3}$ in the sheath due to compression, then drops within the ME (5.0 $cm^{-3}$) and remains nearly steady post-ICME (4.9 $cm^{-3}$). The cooling events, however, start with a slightly higher pre-ICME density (5.0 $cm^{-3}$), see a much larger increase in the sheath (9.4 $cm^{-3}$), and retain a moderately higher density inside the ME (5.6 $cm^{-3}$) and post-ICME (5.7 $cm^{-3}$), consistent with weaker expansion and stronger compression overall (Figure \ref{fig:sea_heating_cooling}).

Furthermore, $B$ remains high in the sheath of both categories, 10.2 nT for Heating and 8.3 nT for Cooling MEs, which could be due to compression but falls within the ME (Figure \ref{fig:sea_heating_cooling}e). However, the Cooling MEs exhibit a slightly higher median field strength (8.5 nT) inside the ME than the Heating ones (8.3 nT), which supports the idea of less expansion and greater field preservation. Heating MEs have a consistently higher flow speed $V$ throughout the event, starting from 423 km s$^{-1}$ and peaking at 505 km s$^{-1}$ in the sheath. Inside the MEs region, the flow slows slightly to 454 km s$^{-1}$ and has an expected expansion-driven decreasing trend, with a further drop to 448 km s$^{-1}$ in the post-region. Cooling MEs begin at a lower speed of 377 km s$^{-1}$, rise to 437 km s$^{-1}$ in the sheath, and fall to 410 km s$^{-1}$ in the ME and post regions. This suggests Heating MEs experience stronger expansion, whereas Cooling MEs retain a more compact, slower-moving structure. Noticeably, the $\beta$ profile in the Heating and Cooling ME categories reveals a common pattern inside the MEs and post-ICME regions but diverges notably in the pre-ICME and sheath regions. In Heating MEs, $\beta$ remains around 1.05 before and after the event, dips to 0.94 in the sheath, and drops further to 0.48 within the ME, reflecting the expected dominance of magnetic pressure in the ejecta and a return to typical solar wind conditions afterward. Cooling MEs, on the other hand, exhibit a sharper decline in $\beta$ inside the ME (0.5) from higher values in both the pre-ICME (1.3) and post-ICME (1.1) regions. However, the key distinction lies in the upstream and sheath regions: Cooling MEs are preceded by less magnetically dominated plasma, consistent with their weaker magnetic fields observed in those regions (Figure \ref{fig:sea_heating_cooling}).

P$_{dyn}$ trends mirror the velocity evolution, peaking in the sheath for both categories, 4.0 nPa in Heating and 3.6 nPa in Cooling, then declining within the MEs and post-ICMEs regions. Notably, Heating MEs maintain similar post-ICME dynamic pressure (2.0 nPa) to Cooling MEs (2.0 nPa) despite differences in $V$ and $N_p$. Interestingly, despite the stronger $P_{dyn}$ in Heating MEs, we note that Cooling MEs tend to produce a more substantial spike in Sym-H at the moment of SSC associated with shock arrival, suggesting that factors beyond $P_{dyn}$ such as strength and orientation of magnetic field in the pre-MEs regions also play an important role in the observed SSC. However, carefully examining, we can see that the jump in the Sym-H from the pre-MEs is certainly larger for Heating MEs than Cooling MEs, consistent with the magnetospheric compression related to increased dynamic pressure at the arrival of shock ahead of MEs (Figure \ref{fig:sea_heating_cooling}).

Furthermore, the Sym-H index declines as the sheath and ME pass (Figure \ref{fig:sea_heating_cooling}h). The rate of this decline is notably steeper in the sheath region for Heating MEs compared to Cooling MEs, suggesting a more rapid onset of magnetic reconnection and geomagnetic disturbance during the early phase of storm development. Furthermore, Heating MEs exhibit lower Sym-H minima within the MEs region, reflecting their tendency to drive a stronger GS. This is consistent with the elevated magnetic field strengths observed in the sheath and front portion of MEs for Heating MEs. During the recovery phase, when the MEs exit and solar wind conditions begin to normalize, both heating and Cooling MEs show a similar pattern towards the recovery phase. This suggests that the post-ICME solar wind environment is almost the same, regardless of the heating or cooling behavior of the preceding MEs. Notably, pre-ICME regions associated with Heating MEs show stronger negative Sym-H values compared to Cooling MEs. This implies a more disturbed magnetospheric environment even before the Heating MEs structure, potentially due to other preceding solar wind structures. Taken together, the stronger $P_{dyn}$, deeper Sym-H depressions, and more pronounced pre- and post-event activity are noted for Heating MEs than Cooling MEs.

We also note that Heating MEs have a slightly higher $V_{exp}$ of 24 km s$^{-1}$ while Cooling MEs expand slowly at just 19 km s$^{-1}$. Although the contrast is not strong, this indicates that Heating MEs undergoing stronger radial expansion are leading to cooling suppression and energy retention within the ME, which is also found for CMEs close to the Sun in the study of \citet{Khuntia2024}. The distortion parameters (DiP-B and DiP-N) for heating and Cooling MEs both hover close to 0.5, indicating relatively symmetric internal structures  (Table \ref{tab:median}). However, Heating MEs show slightly lower DiP values (0.47 for DiP-B and 0.46 for DiP-N), suggesting a mild compression toward the front of the ME. In contrast, Cooling MEs have marginally higher values (both 0.49), pointing to a near-central symmetry with a slight tendency for rear-side compression. This subtle difference hints that Heating MEs may be more dynamically evolving or compressed by the upstream solar wind while Cooling MEs, being more magnetically organized, retain a more balanced structure across their radial span.

\begin{figure*}
    \centering
    \includegraphics[width=1\linewidth]{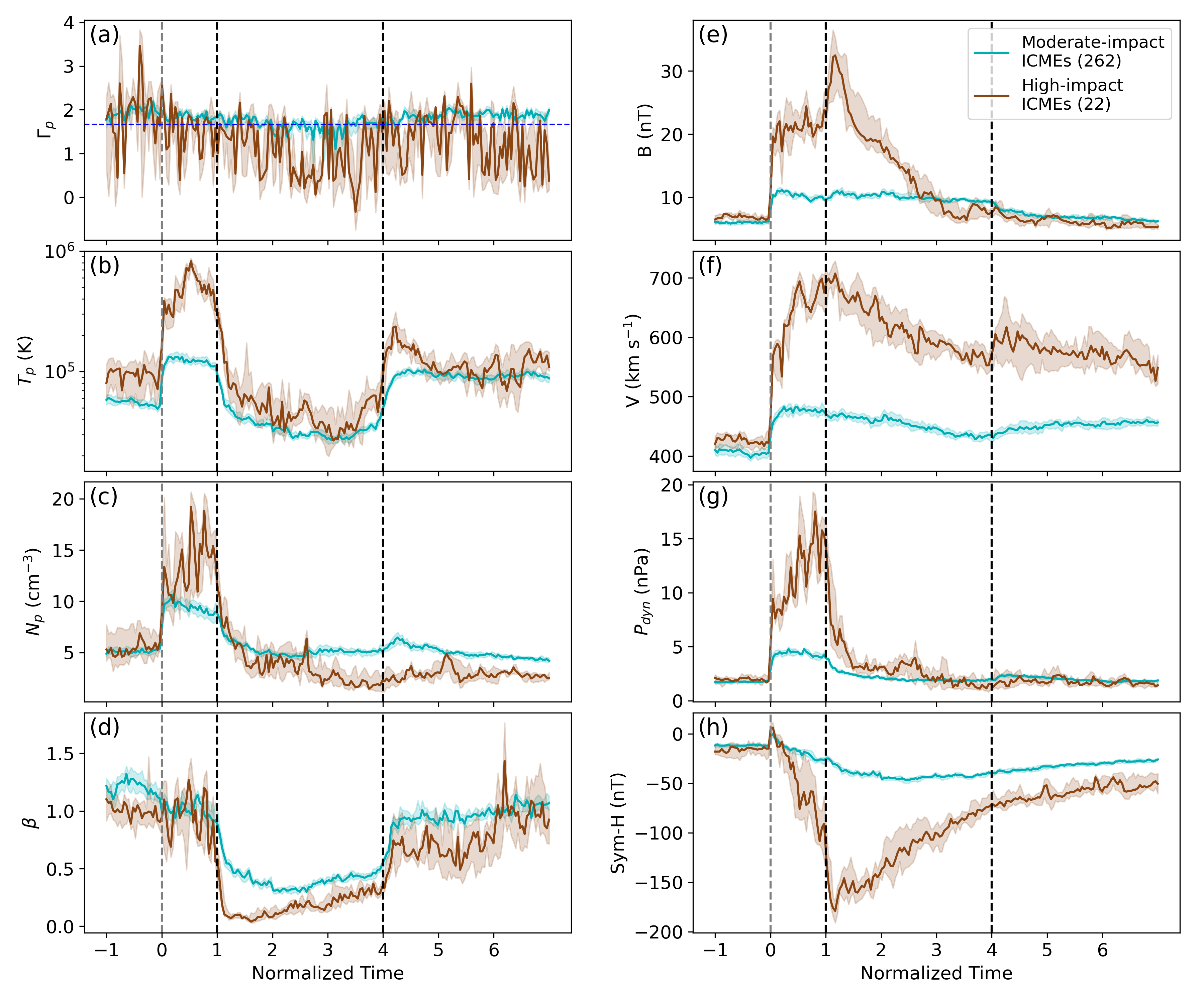}
    \caption{Superposed Epoch Analysis (SEA) showing the median values of parameters across the pre-ICME, sheath, ME, and post-ICME regions: (a) polytropic index ($\Gamma_p$), (b) proton temperature ($T_p$), (c) proton number density ($N_p$), (d) plasma beta ($\beta$), (e) magnetic field strength ($B$), (f) flow speed ($V$), (g) dynamic pressure ($P_{dyn}$), and (h) geomagnetic index (Sym-H). The curves represent: teal for Moderate-impact ICMEs and brown for High-impact ICMEs. Black vertical dashed lines mark the boundaries for the ME region, and gray dashed vertical lines mark the starting sheath region.The background color-shaded regions represent the 68\% (1$\sigma$) confidence intervals obtained via bootstrap resampling.}
    \label{fig:sea_storm}
\end{figure*}

Overall, Heating and Cooling MEs exhibit fundamentally distinct thermodynamic and plasma signatures. Heating MEs are marked by very low $\Gamma_p$, stronger compression, higher speeds, enhanced expansion, and deeper Sym-H depressions, making them more geoeffective drivers of geomagnetic storms. In contrast, Cooling MEs evolve with consistently high $\Gamma_p$, weaker expansion, and more uniform plasma behavior, reflecting enhanced cooling and greater magnetic flux preservation. These contrasting behaviors highlight the importance of thermal state in shaping ICME–magnetosphere interactions and motivate further investigation into how such plasma evolution varies across geomagnetic storm classes.

\subsection{Variation in Plasma Evolution Across Different Storms Classes}\label{subsec:dsea_storm}

To better understand the geoeffectiveness of ICMEs, we classified events into two categories based on their geomagnetic impacts: (i) High-impact ICMEs driving storms with Sym-H $<$ –200 nT (ii) Moderate-impact ICMEs driving storms with –200 nT $< $Sym-H $< $–50 nT. We intentionally excluded events with Sym-H $>$ –50 nT, as these weak cases may dilute the median profiles and obscure meaningful trends in the plasma and magnetic field parameters. This classification is motivated to highlight how variations in solar wind and ICMEs properties, particularly in thermal and magnetic parameters, can be connected to the intensity of storms.

Our analysis indicates a noticeable distinction in thermodynamic behavior ($\Gamma_p$) of High-impact and Moderate-impact ICMEs (Figure \ref{fig:sea_storm}a). In High-impact ICMEs, $\Gamma_p$ drops significantly from 1.97 in the pre-ICME region to 1.56 in the sheath and even further to 0.59 within the MEs before moderately rising to 1.21 in the post-ICME region (Table \ref{tab:median}). This steep decline during the MEs reflects enhanced heating processes within the magnetic structure. Noticeably, there are a lot of variations in $\Gamma_p$ for High-impact ICMEs, indicating localized heating. For Moderate-impact ICMEs, $\Gamma_p$ also decreases across the ICME, from 1.98 (pre-ICME) to 1.79 (sheath), but remains consistently higher than the adiabatic index and compared to High-impact ICMEs. The ME region shows $\Gamma_p$ = 1.69 for Moderate-impact ICMEs, indicative of an adiabatic expansion. The post-ICME value of 1.88 reflects an enhanced cooling than adiabatic conditions. This contrast hints at somewhat more pronounced thermodynamic changes, perhaps due to stronger compression and greater magnetic energy dissipation, in High-impact ICMEs compared to their Moderate-impact counterparts.

\begin{table*}
\caption{Median value of plasma and magnetic field parameters of various regions associated with different categories of ME.}
\label{tab:median}
\centering
%\hfill
\begin{tabular}{llcccccccccc}
\hline
%\hline
Category & Regions & $\Gamma_p$ & $T_p$ & $N_p$ & $V$ & $\beta$ & $B$ & P$_{dyn}$ & DiP-B & DiP-N & $V_{exp}$ \\
 &  &  & ($10^4$ K) & (cm$^{-3}$) & (km s$^{-1}$) &  & (nT) & (nPa) & & & (km s$^{-1}$) \\
\hline
\multirow{4}{3cm}{All ICME}
 & Pre & 2.02 & 5.3 & 5.0 & 400 & 1.2 & 8.6 & 1.7 & - & - & - \\
 & Sheath & 1.83 & 11.5 & 8.6 & 469 & 1.1 & 9.3 & 3.8 & - & - & -  \\
 & ME & 1.74 & 3.0 & 5.3 & 433 & 0.5 & 8.5 & 2.0 & 0.48 & 0.48 & 21\\
 & Post & 1.91 & 7.6 & 5.4 & 428 & 1.1 & 6.6 & 2.0 & - & - & - \\
 \hline
\multirow{4}{3cm}{MC}
 & Pre & 2.09 & 4.5 & 6.6 & 375 & 1.4 & 5.6 & 1.6 & -  & -  &  -  \\
 & Sheath & 1.83 & 10.1 & 11.4 & 455 & 1.1 & 10.4 & 4.9 & -  & -  & -   \\
 & ME & 1.75 & 2.8 & 6.3 & 421 & 0.3 & 11.3 & 2.2 & 0.48 & 0.46 &  21\\
 & Post & 1.88 & 7.5 & 5.7 & 432 & 1.1 & 6.8 & 2.1 & -  &  - &  - \\
 \hline
\multirow{4}{3cm}{Non-MC} 
 & Pre & 1.95 & 6.3 & 4.8 & 417 & 1.1 & 6.1 & 1.7 & -  & -  & -   \\
 & Sheath & 1.83 & 12.8 & 7.2 & 479 & 1.1 & 8.7 & 3.4 &  - & -  & -  \\
 & ME & 1.72 & 3.1 & 5.0 & 438 & 0.7 & 7.1 & 1.9 & 0.48 & 0.49 &  20\\
 & Post & 1.93 & 7.7 & 5.2 & 428 & 1.1 & 6.6 & 1.9 & -  & -  & -  \\
 \hline
\multirow{4}{2.5cm}{Heating MEs}  
& Pre   & 1.84 & 6.2 & 5.0 & 423 & 1.0 & 6.6 & 1.8 & - & - & - \\
& Sheath & 1.64 & 15.3 & 7.7 & 505 & 0.9 & 10.2 & 4.0 & - & - & - \\
& ME     & 0.42 & 3.1 & 5.0 & 454 & 0.5 & 8.3 & 2.2 & 0.47 & 0.46 & 24 \\
& Post   & 1.59 & 8.6 & 4.9 & 448 & 1.0 & 6.7 & 2.0 & - & - & - \\
\hline
\multirow{4}{2.5cm}{Cooling MEs}  
& Pre   & 2.16 & 4.9 & 5.0 & 377 & 1.3 & 5.5 & 1.6 & - & - & - \\
& Sheath & 1.99 & 9.3 & 9.4 & 437 & 1.2 & 8.3 & 3.7 & - & - & - \\
& ME     & 2.07 & 2.9 & 5.6 & 412 & 0.5 & 8.5 & 2.0 & 0.49 & 0.49 & 19 \\
& Post   & 2.14 & 6.8 & 5.7 & 410 & 1.1 & 6.6 & 2.0 & - & - & - \\
\hline
\multirow{4}{2.5cm}{High-impact ICMEs}
 & Pre & 1.97 & 9.1 & 7.3 & 422 & 1.1 & 6.8 & 1.8 & - & - & - \\
 & Sheath & 1.56 & 58.4 & 17.2 & 676 & 1.0 & 22.8 & 14.5 & - & - & - \\
 & ME & 0.59 & 4.5 & 3.2 & 615 & 0.1 & 13.4 & 3.0 & 0.30 & 0.38 & 70 \\
 & Post & 1.21 & 10.8 & 3.1 & 573 & 0.8 & 6.2 & 2.0 & - & - & - \\
 \hline
\multirow{4}{3cm}{Moderate-impact ICMEs}
 & Pre & 1.98 & 4.9 & 5.4 & 409 & 1.1 & 6.2 & 1.8 & - & - & - \\
 & Sheath & 1.79 & 13.1 & 9.7 & 479 & 1.0 & 10.7 & 4.7 & - & - & -\\
 & ME & 1.69 & 3.3 & 5.2 & 455 & 0.4 & 10.0 & 2.1 & 0.46 & 0.45 & 22 \\
 & Post & 1.88 & 9.1 & 5.0 & 456 & 1.0 & 7.0 & 2.0 & - & - & - \\
 \hline
\end{tabular}
%\hfill
\end{table*}

The comparison between ME part of High-impact and Moderate-impact ICMEs reveals statistically significant differences in the thermodynamic behavior as characterized by $\Gamma_p$, although the effect size remains modest. The KS test yields a test statistic of $D = 0.106$ with a p-value of $9.52 \times 10^{-98}$, indicating a significant difference in the shape of the two $\Gamma_p$ distributions. The Mann–Whitney U test supports this finding, with a p-value of $1.09 \times 10^{-89}$, suggesting a meaningful shift in central tendency. Cliff’s Delta of $-0.116$ indicates a small effect size, showing that a randomly selected moderate-impact ICME has a higher $\Gamma_p$ than a high-impact ICME in approximately 56\% of comparisons. This is consistent with the observed median difference of $\Delta \Gamma_p = -0.56$ between the two populations, which is statistically significant according to the permutation test ($p < 10^{-5}$). The KDE plot (Figure~\ref{fig:kde}c) illustrates this subtle but consistent separation, with moderate-impact events exhibiting a denser concentration in the higher-$\Gamma_p$ regime. This trend is further reflected in the SEA: during the main storm phase, the proportion of time bins with $\Gamma_p < 1.67$ is substantially higher for High-impact events (66\%) compared to Moderate-impact events (19\%), suggesting more prevalent heating in stronger storms. Conversely, Moderate-impact ICMEs maintain $\Gamma_p > 1.67$ in 81\% of time bins, indicating more adiabatic behavior. Taken together, these results indicate that High-impact ICMEs are statistically more likely to exhibit lower polytropic indices, which may be associated with stronger energy dissipation or internal heating processes contributing to their geoeffectiveness.

Noticeably, $T_p$ exhibits a highly elevated peak in the sheath region for High-impact ICMEs, reaching a median of $58.4 \times 10^4 K$, compared to $13.1 \times 10^4 K$ in Moderate-impact ICMEs (Table \ref{tab:median}). This dramatic heating in strong events points to intense shock compression region ahead of MEs. In the pre-ICME region, $T_p$ is also higher in High-impact ICMEs ($9.1 \times 10^4 K$) than in Moderate-impact ones ($5.0 \times 10^4 K$), suggesting a hotter upstream environment. Inside the ME, the temperature decreases substantially in both cases, to $4.5 \times 10^4 K$ (High-impact) and $3.3 \times 10^4 K$ (Moderate-impact), consistent with the expected expansion-related cooling of the magnetically dominated ejecta. The post-ICME region shows similarly elevated levels as $10.8 \times 10^4 K$ for High-impact vs. $9.1 \times 10^4 K$ for Moderate-impact ICMEs. Moreover, $N_p$ follows an expected pattern of enhancement in the sheath region and dilution inside the MEs, but with key differences in magnitude. In High-impact MEs, the sheath region shows a sharp rise to 17.1 $cm^{-3}$ from 7.3 $cm^{-3}$ in the pre-ME region, followed by a significant drop to 3.1 $cm^{-3}$ within the MEs and remaining low at 3.1 $cm^{-3}$ post-MEs. In contrast, Moderate-impact ICMEs exhibit a less dense sheath (9.7 $cm^{-3}$) and maintain higher density in MEs (5.2 $cm^{-3}$) and post-MEs density (5.0 $cm^{-3}$). This indicates that strong events are accompanied by more efficient plasma compression in the sheath but more depleted interiors, consistent with stronger expansion effects.

Notably, $\beta$ clearly differentiates the magnetic dominance in ME regions between the drivers of two storm categories (Figure \ref{fig:sea_storm}d). For High-impact ICMEs, $\beta$ plunges from 1.1 (pre-ICME region) and 1.0 (sheath) to a remarkably low 0.1 inside the MEs before rising to 0.8 post-ICME region. Moderate-impact ICMEs follow a similar trend but with much milder variation, maintaining values of 1.1 (pre-ICME), 1.0 (sheath), 0.4 (ME), and 1.0 (post-ICME). The significantly lower $\beta$ within MEs of High-impact ICMEs reflects stronger magnetic pressure and flux content, contributing directly to enhanced storm-driving potential. We found that $B$ is markedly higher in High-impact ICMEs, peaking at 22.8 nT in the sheath and 13.4 nT in the MEs, compared to 10.7 nT in the sheath and 10.1 nT in MEs driving Moderate-impact ICMEs (Figure \ref{fig:sea_storm}e). Interestingly, inside the MEs of High-impact ICMEs, $B$ decreases rapidly from Leading to the trailing edge of MEs. This is also evident in the DiP-B parameters, which have a value of 0.30, indicating stronger magnetic compression close to the leading part of ME (Table \ref{tab:median}). However, the DiP-N=0.38 suggests a density compression but as front-loaded as magnetic compression. This asymmetry in High-impact ICMEs is likely a consequence of stronger dynamic interactions with the ambient solar wind and enhanced expansion, leading to sharper, more compressed fronts and more depleted trailing regions. DiP-B=0.46 and DiP-N=0.45 suggest a more symmetrical magnetic field and density inside ME of Moderate-impact ICMEs than the High-impact ones. These elevated magnetic field strengths in the sheath and MEs are crucial for the development of strong GS, particularly when associated with sustained southward components.

As expected, $V$ is significantly higher in High-impact ICMEs, indicating faster propagating ICMEs (Figure \ref{fig:sea_storm}f). $V$ increases from 422 km s$^{-1}$ (pre) to 676 km s$^{-1}$ in the sheath, peaks at 615 km s$^{-1}$ inside the ME, and remains elevated at 573 km s$^{-1}$ post-ICME. Moderate-impact ICMEs, by contrast, feature more subdued speeds: 409 km s$^{-1}$ (pre), 479 km s$^{-1}$ (sheath), 455 km s$^{-1}$ (ME), and 456 km s$^{-1}$ (post). The sharp speed enhancements in High-impact ICMEs are indicative of greater kinetic energy and geoeffectiveness, which are often associated with fast CMEs. Notably, $V_{exp}$ is substantially greater in High-impact ICMEs, with a median value of 70 km s$^{-1}$ compared to 22 km s$^{-1}$ in Moderate-impact ICMEs. This increased expansion suggests a higher internal pressure and a dynamically evolving structure. We found that $P_{dyn}$ shows an extreme increase during the sheath phase of High-impact ICMEs, reaching 14.5 nPa, an increase from the pre-ICME level of 1.8 nPa. It then drops to 3.0 nPa in the ME and further to 2.0 nPa in post-ICME region. Moderate-impact ICMEs also show a sheath enhancement (4.7 nPa from 1.8 nPa in pre-ICME), but the ME and post-ICME values (2.1 and 2.0 nPa, respectively) are more stable. This difference indicates that the initial solar wind compression of the magnetosphere due to High-impact ICMEs is much more intense than Moderate-impact ICMEs.

The Sym-H index, which reflects geomagnetic response, displays an obviously distinct behavior between High-impact and Moderate-impact ICMEs, particularly around and after storm onset (Figure \ref{fig:sea_storm}h). In the pre-ICME region, both categories of ICMEs exhibit nearly identical Sym-H levels, indicating similar quiet geomagnetic conditions prior to the arrival of the disturbances. Following this, both types of ICMEs show a comparable sudden positive spike in Sym-H (SSC) despite the significantly stronger $P_{dyn}$ associated with the shock driven by High-impact ICMEs. This suggests that SSC magnitude may depend on other coupling conditions beyond $P_{dyn}$ alone. A sharp drop in Sym-H follows in the sheath region, where High-impact ICMEs show a much steeper and deeper descent, indicating stronger magnetic reconnection and more effective energy transfer to the magnetosphere. Notably, the lowest Sym-H value, marking the main phase of the storm, occurs near the leading edge of the MEs for High-impact ICMEs. Notably, the recovery phase of the storms caused by High-impact ICMEs typically begins shortly after the leading edge of the ME and extends throughout the remaining ME and post-ICME region.

Although the geomagnetic response is primarily driven by $B_z$ and $P_{dyn}$ \citep{Echer2008, Richardson2011, Zhao2021}, our results show that these parameters are intrinsically connected to the thermodynamic state of the ICME, as described by $\Gamma_p$. The lower $\Gamma_p$ values observed in High-impact ICMEs imply enhanced heat retention or ongoing energy input, which maintains higher internal pressure and magnetic tension within the ejecta. Such over-pressured MEs behave as stronger pistons, compressing the upstream solar wind more effectively and generating intense sheath regions with amplified magnetic fields. Through field-line draping and flux pile-up, this compression can locally enhance and rotate $B_z$ \citep{Manchester2004, Manchester2017}, thereby strengthening magnetospheric coupling and deepening the storm main phase. In contrast, Moderate-impact ICMEs, characterized by higher and more adiabatic $\Gamma_p$, expand more freely, resulting in weaker sheath compression, reduced magnetic field amplification, and lower dynamic pressure at 1 AU. Hence, the thermodynamic state of the ejecta can regulate how kinetic and magnetic energy are partitioned and transferred to the magnetosphere. The observed relationship between low $\Gamma_p$, high internal pressure, and enhanced magnetic compression provides a thermodynamic basis for the greater geoeffectiveness of High-impact ICMEs. Future event-specific studies will aim to quantify the relative contributions of these heating-related processes to magnetic field amplification and storm intensity, thereby establishing a more direct link between CME thermal evolution and geoeffective potential.

\section{Conclusion} \label{sec:discussion}

This study provides a comprehensive thermal characterization of MEs and associated plasma properties across Solar Cycles 23, 24, and the rising phase of 25, based on the behavior of the polytropic index ($\Gamma_p$). By categorizing MEs into Heating and Cooling types, and applying a refined classification scheme (Major-heating, Isothermal, Adiabatic, Major-cooling), we uncover critical insights into their solar-cycle dependence, thermodynamic evolution, plasma signatures, and geoeffectiveness. The main conclusions are as follows:

\begin{enumerate}
    
    \item Heating MEs dominate during solar maxima and descending phases (especially in SC23), reflecting strong thermal energy input during eruption and/or interplanetary propagation. In contrast, the rising phases of SC23, SC24, and SC25 show a predominance of Cooling MEs, suggesting reduced heating efficiency or different source-region conditions.

    \item A progressive increase in median $\Gamma_p$ values from 1.46 in SC23 to 1.87 in SC24 indicates a shift from heating to cooling-dominated thermal states in MEs. The fraction of heating MEs decreases from approximately 59\% in SC23 to 34\% in SC24, suggesting that this change could arise either from inherent differences in the MEs near the Sun or from their altered evolution during transit.

    \item The refined classification using $\Gamma_p$ (e.g., Major-heating, Isothermal, Adiabatic, Major-cooling) shows that MEs rarely behave adiabatically or isothermally; they are thermodynamically active systems. The majority of MEs show extreme values of $\Gamma_p$ ($< $0.9 or $> $1.77), highlighting intense energy exchange, either strong energy gain or significant thermal loss.

    \item Heating MEs exhibit strong solar cycle modulation in $\Gamma_p$, $T_p$, and $V_{exp}$, indicating active and variable heating processes tied to solar activity levels and eruption energetics. Their $\Gamma_p$ shows large year-to-year variation with distinct lower values (strong heating) near solar maxima and minima phases, especially in SC23 and SC24. Despite low in-situ $T_p$ in some years (e.g., 2013), high $V_{exp}$ and low $\Gamma_p$ indicate significant in-transit heating, suggesting that initial eruption condition and evolution en route jointly determine ME thermal states.

    \item Cooling MEs remain thermodynamically stable and consistent across cycles, with nearly constant $\Gamma_p\approx2$, suggesting enhanced cooling beyond what is expected from natural adiabatic expansion. Interestingly, despite their cooling behavior, Cooling MEs exhibit $T_p$ values comparable to those of Heating MEs, suggesting that they either originate with higher internal energy or retain substantial thermal energy during their interplanetary propagation before reaching 1 AU.

    \item More geoeffective ICMEs are typically MCs, characterized by strong magnetic fields, low plasma beta, and distinct thermal properties such as low proton temperatures and $\Gamma_p$ $< $ 1.67. Their coherent magnetic structure and dominant magnetic pressure often lead to deeper Sym-H depressions compared to those linked with non-MC ICMEs.

    \item Heating and Cooling MEs evolve through fundamentally different thermodynamic pathways: Heating MEs exhibit pronounced non-adiabatic behavior, characterized by very low $\Gamma_p$, along with elevated dynamic and magnetic pressures, significant internal energy deposition, and enhanced geoeffectiveness. Moreover, Heating MEs exhibit front-loaded compression, asymmetry in internal density structure (low DiP), and faster expansion speeds. In contrast, Cooling MEs are found to exceed adiabatic expectations (high $\Gamma_p$), maintain magnetic coherence, and result in milder geomagnetic responses.

    \item High-impact ICMEs are typically Heating MEs (primarily MCs) ($\Gamma_p = 0.59$), showing stronger magnetic fields, low plasma beta, higher bulk and expansion speeds, and stronger dynamic compression in the sheath region. The thermal state of MEs, alongside their other geoeffective properties, may be a good indicator of the ICME’s potential to drive strong geomagnetic storms. Importantly, stronger storms caused by High-impact ICMEs often result not only from the properties of the ME itself but also from significant contributions by the preceding sheath structures.

\end{enumerate}

Overall, our findings highlight the importance of incorporating the thermal state of ICMEs, alongside conventional plasma parameters, to gain deeper insights into their interplanetary evolution, classification, and potential geoeffectiveness at 1~AU. The $\Gamma_p$-based framework, and its statistical characterization at 1~AU, offers a novel perspective on CME energetics and provides a useful diagnostic for probing the underlying thermodynamic processes. However, such thermodynamic diagnostics (e.g., $\Gamma_p$) should not be interpreted in isolation as direct indicators of geoeffectiveness; rather, they must be assessed in conjunction with other plasma and magnetic field parameters. Only through such combined diagnostics can the role of CME thermal evolution in shaping geomagnetic responses be meaningfully constrained. Looking forward, this approach may support the development of improved space weather forecasting schemes and motivate further investigations into the physical drivers of CME heating and cooling in the heliosphere.

\section*{Acknowledgements}

We appreciate the anonymous referee for the constructive comments and valuable suggestions. We acknowledge the use of data from the OMNI solar-wind database, compiled by the Space Physics Data Facility at NASA Goddard Space Flight Center. 

%%%%%%%%%%%%%%%%%%%%%%%%%%%%%%%%%%%%%%%%%%%%%%%%%%
\section*{Data Availability}

All the observational input data sets used in this study are publicly available at NASA Coordinated Data Analysis Web (CDAWeb; \url{https://cdaweb.gsfc.nasa.gov/}).

%%%%%%%%%%%%%%%%%%%% REFERENCES %%%%%%%%%%%%%%%%%%

%%%%%%%%%%%%%%%%% APPENDICES %%%%%%%%%%%%%%%%%%%%%

\appendix

\section{List of all studied ICMEs}\label{App_sec:list}

\begin{table*}
%\small
\centering
\caption{List of ICMEs studied from June 1996 to December 2024, covering solar cycles 23 to the rising phase of 25. The table includes the corresponding derived median value of the polytropic index ($\Gamma_p$), median value of the uncertainty in $\Gamma_p$ ($\Delta\Gamma_p$) due to fractional measurement uncertainties in proton density ($\delta N_p=$ 2\%) and proton temperature ($\delta T_p=$5\%), expansion speed ($V_{exp}$), and distortion parameters for magnetic field (DiP-B) and proton number density (DiP-N). This table is available in its entirety in machine-readable format. A portion is shown here for illustrative purposes, indicating the form and content. The date time format used in columns 2–4 is \textit{YYYY-MM-DD HH:MM}.}
\label{tab:ICME_list}
 
\begin{tabular}{rccccccccc}
\hline
\textbf{No. }& \textbf{Disturbance time} & \textbf{ME start time} & \textbf{ME end time} & \textbf{ME type}  & \textbf{Median $\Gamma_p$} & \textbf{Median $\Delta \Gamma_p$} & \textbf{$V_{exp}$} &\textbf{ DiP-B} & \textbf{DiP-N} \\
 & &  &  &  &  & & ($\mathrm{km\ s^{-1}}$) &  &  \\
\hline
1 & 1996-07-01 13:20:00 & 1996-07-01 18:00:00 & 1996-07-02 11:00:00 & 2 & 2.459 & 0.57 & 12 & 0.522 & 0.525 \\
2 & 1996-08-07 06:00:00 & 1996-08-07 12:00:00 & 1996-08-08 10:00:00 & 2 & 2.23 & 1.135 & 10 & 0.578 & 0.439 \\
3 & 1996-12-23 16:00:00 & 1996-12-23 17:00:00 & 1996-12-25 11:00:00 & 2 & 2.032 & 0.725 & 53 & 0.589 & 0.515 \\
4 & 1997-01-10 01:04:00 & 1997-01-10 04:00:00 & 1997-01-11 02:00:00 & 2 & 0.777 & 0.333 & 44 & 0.501 & 0.777 \\
5 & 1997-02-09 13:21:00 & 1997-02-10 02:00:00 & 1997-02-10 19:00:00 & 2 & 2.117 & 0.46 & 32 & 0.52 & 0.195 \\
6 & 1997-04-10 17:45:00 & 1997-04-11 06:00:00 & 1997-04-11 19:00:00 & 2 & 1.342 & 0.23 & -7 & 0.467 & 0.521 \\
7 & 1997-04-21 06:00:00 & 1997-04-21 10:00:00 & 1997-04-23 04:00:00 & 2 & 2.219 & 0.638 & 31 & 0.512 & 0.444 \\
8 & 1997-05-15 01:59:00 & 1997-05-15 09:00:00 & 1997-05-16 00:00:00 & 2 & 1.967 & 0.395 & -12 & 0.443 & 0.4 \\
9 & 1997-05-26 09:57:00 & 1997-05-26 16:00:00 & 1997-05-27 10:00:00 & 2 & 1.936 & 0.564 & 11 & 0.495 & 0.443 \\
10 & 1997-06-08 16:36:00 & 1997-06-08 18:00:00 & 1997-06-10 00:00:00 & 2 & 2.039 & 0.517 & 19 & 0.538 & 0.402 \\
\hline
\end{tabular}
\end{table*}

\section{Annual distribution of magnetic ejecta with various thermal states at 1 AU} \label{App_sec:thrmal}

\begin{table*}
%\small
\centering
\caption{Annual distribution of magnetic ejecta with various thermal states at 1 AU.}
\label{tab:thermal_state_perce}
%\begin{adjustbox}{height=0.3\textheight} 
\begin{tabular}{cccccc}
\hline
\textbf{Year} & 
\textbf{Major-heating} & 
\textbf{Isothermal} & 
\textbf{Heating} & 
\textbf{Adiabatic} & 
\textbf{Major-Cooling} \\
& \textbf{Count (\%)} & \textbf{Count (\%)} & \textbf{Count (\%)} & \textbf{Count (\%)} & \textbf{Count (\%)} \\
\hline
1996 & 0 (0) & 0 (0) & 0 (0) & 0 (0) & 3 (100) \\
1997 & 1 (5) & 0 (0) & 1 (5) & 1 (5) & 19 (86) \\
1998 & 5 (14) & 1 (3) & 3 (8) & 4 (11) & 23 (64) \\
1999 & 11 (33) & 0 (0) & 2 (6) & 6 (18) & 14 (42) \\
2000 & 24 (48) & 0 (0) & 5 (10) & 5 (10) & 16 (32) \\
2001 & 29 (60) & 0 (0) & 9 (19) & 2 (4) & 8 (17) \\
2002 & 12 (46) & 0 (0) & 4 (15) & 4 (15) & 6 (23) \\
2003 & 9 (43) & 0 (0) & 2 (9) & 3 (14) & 7 (33) \\
2004 & 9 (43) & 0 (0) & 2 (10) & 4 (19) & 6 (29) \\
2005 & 20 (65) & 0 (0) & 7 (23) & 2 (6) & 2 (6) \\
2006 & 6 (46) & 1 (8) & 1 (8) & 2 (15) & 3 (23) \\
2007 & 1 (50) & 0 (0) & 0 (0) & 0 (0) & 1 (50) \\
2008 & 0 (0) & 0 (0) & 0 (0) & 1 (33) & 2 (67) \\
2009 & 2 (18) & 0 (0) & 1 (9) & 3 (27) & 5 (45) \\
2010 & 3 (20) & 0 (0) & 1 (7) & 1 (7) & 10 (67) \\
2011 & 5 (16) & 0 (0) & 2 (6) & 5 (16) & 20 (63) \\
2012 & 3 (9) & 0 (0) & 2 (6) & 4 (12) & 26 (74) \\
2013 & 8 (33) & 0 (0) & 0 (0) & 3 (13) & 13 (54) \\
2014 & 6 (30) & 0 (0) & 1 (5) & 4 (20) & 9 (45) \\
2015 & 10 (33) & 0 (0) & 4 (13) & 3 (10) & 13 (43) \\
2016 & 5 (38) & 0 (0) & 1 (8) & 1 (8) & 6 (46) \\
2017 & 1 (11) & 0 (0) & 1 (11) & 1 (11) & 6 (67) \\
2018 & 1 (13) & 0 (0) & 2 (25) & 0 (0) & 5 (63) \\
2019 & 1 (14) & 0 (0) & 0 (0) & 0 (0) & 6 (86) \\
2020 & 2 (50) & 0 (0) & 0 (0) & 0 (0) & 2 (50) \\
2021 & 0 (0) & 0 (0) & 0 (0) & 0 (0) & 10 (100) \\
2022 & 1 (5) & 0 (0) & 1 (5) & 1 (5) & 16 (84) \\
2023 & 1 (4) & 0 (0) & 5 (20) & 4 (16) & 15 (60) \\
2024 & 1 (4) & 1 (4) & 5 (19) & 4 (15) & 16 (59) \\
\hline
\end{tabular}
%\end{adjustbox}
\end{table*}

\section{Kernel Density Estimates (KDE) of Polytropic Index for Different ICME Classifications} \label{App_sec:kde}

\begin{figure*}
    \centering
    \includegraphics[width=0.4\linewidth]{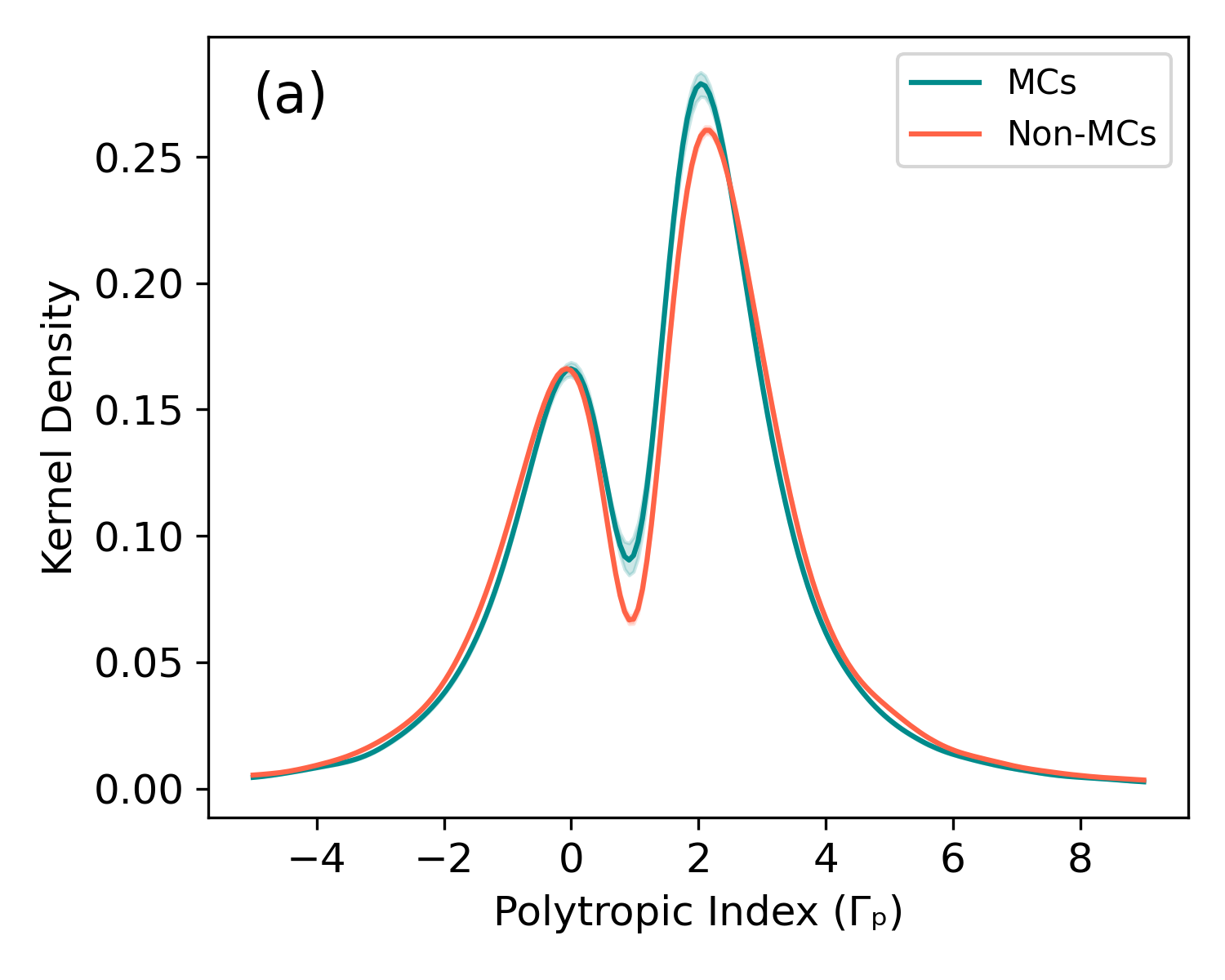}
     \includegraphics[width=0.4\linewidth]{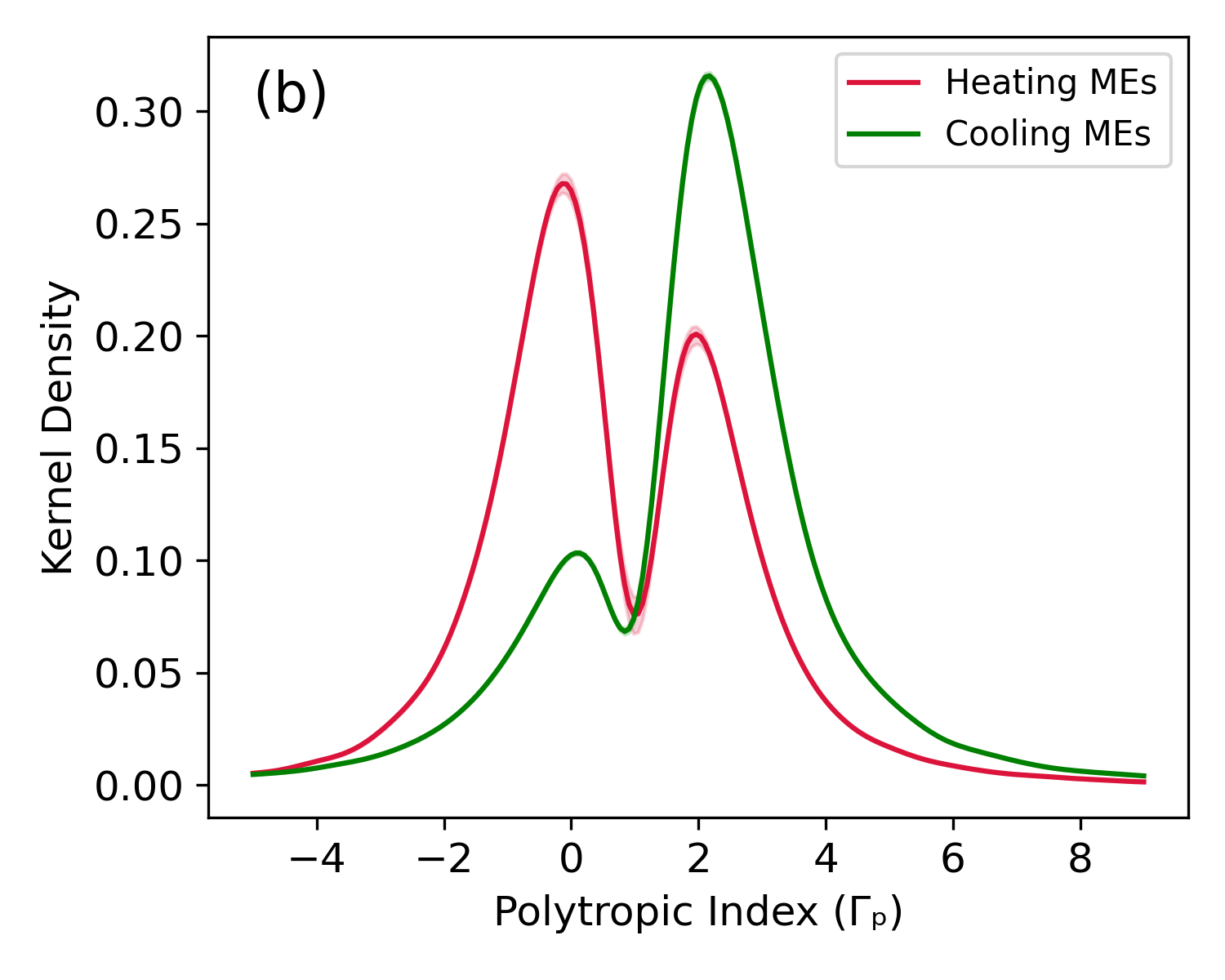}
      \includegraphics[width=0.4\linewidth]{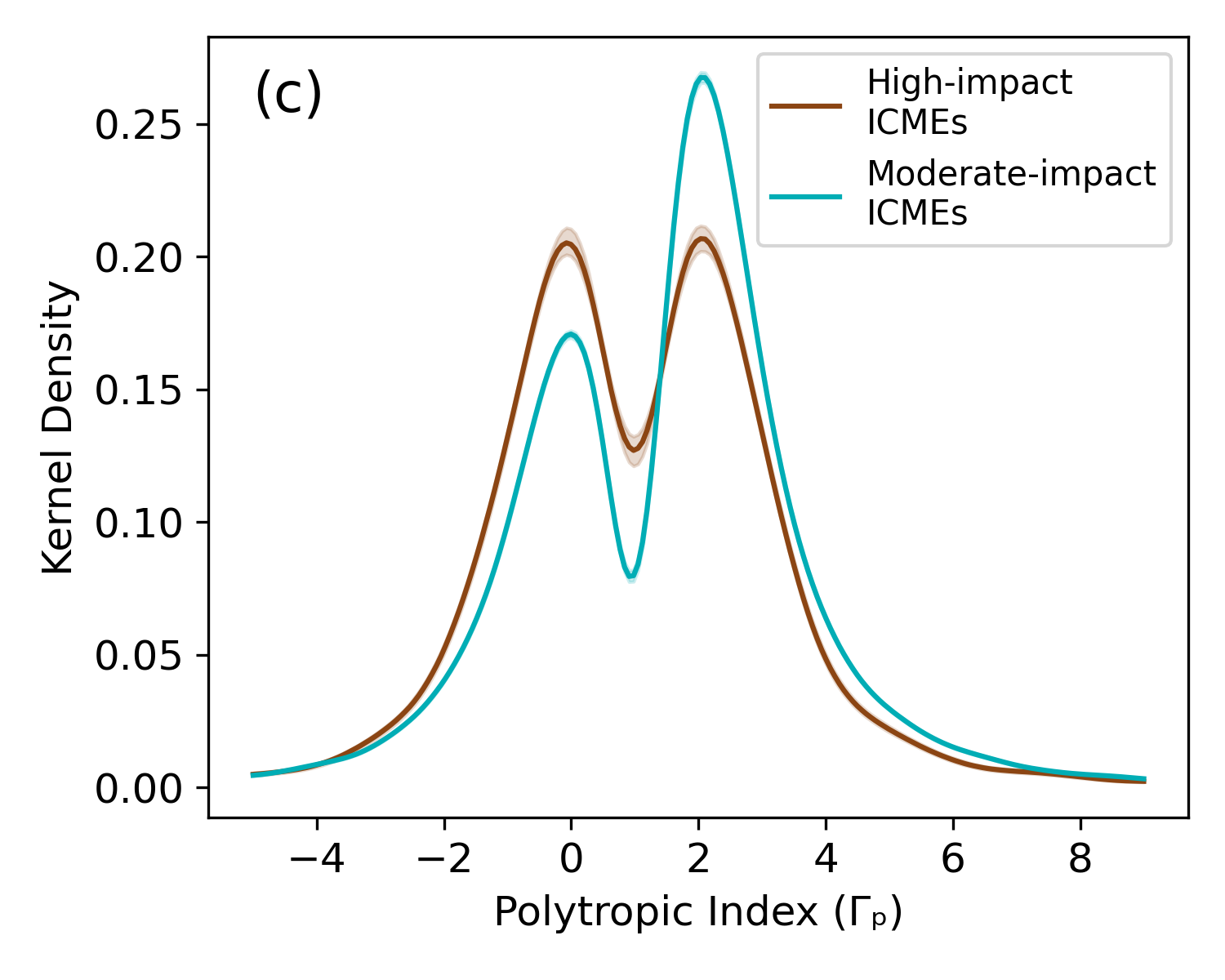}
    \caption{Comparison of $\Gamma_p$ distributions using KDE plots with 68\% bootstrap confidence bands for ME part of (a) MCs vs. Non-MCs, (b) Heating vs. Cooling MEs, and (c) High-impact vs. Moderate-impact ICMEs. These plots highlight differences in thermodynamic properties across ICME subsets.}
    \label{fig:kde}
\end{figure*}

% Don't change these lines
\bsp	% typesetting comment
\label{lastpage}
\end{document}